\documentclass[iop]{emulateapj}

\def\mdot{\hbox{$\dot {\it M}$}}
\def\micron{$\mu$m}

\def\rsun{R_{\sun}}
\def\msun{M_{\sun}}
\def\lsun{L_{\sun}}
\def\arcsec{$^{\prime\prime}$} 
\def\h2{H$_2$}

\def\curf{{\cal F}}
\newcommand\msunyr{\rm {\it M}_{\odot}\,yr^{-1}}
\newcommand\be{\begin{equation}}
\newcommand\en{\end{equation}}

\usepackage{graphicx}
\usepackage[caption=false]{subfig}
\usepackage{xcolor}
\usepackage{url}
\usepackage{hyperref}

\newcounter{column_number}
\begin{document}

\shortauthors{Espaillat et al.}
\shorttitle{X-rays, the \h2 Bump, and Accretion in TTS}

\title{Using Multiwavelength Variability to Explore the Connection between X-ray Emission, the Far-Ultraviolet \h2 Bump, and Accretion in T Tauri Stars}

\author{C. C. Espaillat\altaffilmark{1},
C. Robinson\altaffilmark{1}, 
S. Grant\altaffilmark{1}, \& 
M. Reynolds\altaffilmark{2} 
}

\altaffiltext{1}{Department of Astronomy \& The Institute for Astrophysical Research, Boston University, 725 Commonwealth Avenue, Boston, MA 02215, USA; cce@bu.edu, \hbox{connorr@bu.edu}, sierrag@bu.edu}
\altaffiltext{2}{Department of Astronomy, University of Michigan, 830 Dennison Building, 500 Church
Street, Ann Arbor, MI 48109, USA; markrey@umich.edu} 

\begin{abstract}
	
The high-energy radiation fields of T Tauri stars (TTS) should affect the surrounding circumstellar disk, having implications for disk transport and heating. Yet, observational evidence of the effect of high-energy fields on disks is scarce. Here we investigate the connection between X-ray emission and the innermost gas disk by leveraging the variability of TTS. We obtained multiple epochs of coordinated data (taken either simultaneously or within a few hours) of accreting TTS with the {\it Hubble Space Telescope}, the {\it Neil Gehrels Swift Observatory}, and the {\it Chandra X-ray Observatory}. We measured the far-ultraviolet (FUV) \h2 {bump feature at 1600 \AA}, which traces gas $<1$ AU from the star; the near-ultraviolet (NUV) emission, from which we extract the accretion luminosity; and also the X-ray luminosity. We do not find a correlation between the FUV \h2 bump and X-ray luminosity. Therefore, an observable tracer of the effect of X-ray ionization in the innermost disk remains elusive. We report a correlation between the FUV \h2 bump and accretion luminosity, linking this feature to the disk surface density. We also see a correlation between the X-ray luminosity and the accretion column density, implying that flaring activity may influence accretion. These results stress the importance of coordinated multiwavelength work to understand TTS.

\end{abstract}

\slugcomment{Accepted to ApJ on April 6, 2019.}


\section{Introduction} \label{intro} 

Studying the structure and composition of protoplanetary disks is important in order to understand the initial conditions of planet formation.  In this vein, many studies have probed the dust and gas content of such disks \citep[e.g., reviews by][]{henning13,andrews15}, particularly around low-mass ($<1\msun$) pre-main sequence stars (i.e., T~Tauri stars, TTS).  Studying the interaction between the disk and its variable young star is crucial since the star is the dominant heating source of the disk, which may lead to disk structural and compositional changes. In addition, high-energy radiation from the central star has important implications on the fundamentals of physical transport processes. In particular, X-ray photons can partially ionize and heat the gas in the upper atmosphere of the disk to temperatures up to $\sim4000$--5000~K \citep{glassgold07,meijerink08}.  Therefore, X-ray irradiation especially should play a crucial role in disk ionization, which is important for disk accretion via magnetorotational instability \citep[e.g., see review by][]{hartmann16}.  However, robust observational connections between high-energy stellar radiation fields and circumstellar material remain elusive.

In TTS, X-ray emission is thought to arise predominantly from the stellar corona \citep[i.e., originating in stellar magnetic activity;][]{feigelson02,brickhouse10}. The effect of X-ray irradiation on the disk has been seen observationally, namely through mid-infrared (MIR) forbidden line emission.  [\ion{Ne}{2}] emission lines have been detected in more than 50 TTS \citep{pascucci07, espaillat07a, lahuis07, flaccomio09, gudel10, baldovin11,szulagyi12,espaillat13} and have been attributed to X-ray ionization and heating \citep{glassgold07}, although extreme-UV (EUV) photons may also play a role \citep{hollenbach09, espaillat13}. Recently, variability in X-ray--sensitive millimeter gas lines with the Atacama Large Millimeter Array point to X-ray--driven time-dependent chemistry in the outer disk \citep{cleeves17}. A connection between X-ray emission and the innermost disk, where accretion onto the star occurs and terrestrial planets are formed, remains to be seen. 

One potential tracer of the connection between the X-ray radiation field and the gas in the innermost disk lies within the broad emission feature at 1600~{\AA}. This feature is a combination of Ly$\alpha$-fluoresced \h2 emission lines and broad \h2 continuum emission, the latter commonly referred to as the ``\h2 bump.'' 
The far-ultraviolet (FUV) \h2 bump was first identified by \citet{herczeg04} and \citet{bergin04} in the spectra of classical TTS  \citep[CTTS; i.e., accreting TTS;][]{hartmann16} and has been observed in several disks around CTTS \citep{ingleby09, france17}. In general, FUV \h2 emission traces gas in roughly the innermost $\sim1$ AU of the disk \citep{herczeg02}. \citet{ingleby09} found that CTTS display the FUV \h2 bump while weak-lined TTS (WTTS; i.e., non-accreting stars) do not, linking the \h2 bump to the presence of gas in the inner disk.  The \h2 bump has been proposed to be due to collisional excitation of \h2 by fast electrons in the inner disk \citep{weintraub00,bary02,herczeg04,bergin04}. Collisional excitation occurs when electrons created by X-ray ionization of metals in the inner disk ionize hydrogen and helium and create an abundance of hot electrons.  The electrons collisionally excite \h2, and one de-excitation path produces continuum emission.  However, more recently, it has been suggested that the \h2 bump is powered by Lyman~$\alpha$ (Ly$\alpha$) photons, particularly Ly$\alpha$-driven dissociation of \h2O in the inner disk \citep{france17}. Excitation by Ly$\alpha$ photons will populate the upper levels of \h2, and a fluorescent spectrum will be emitted as it de-excites. \citet{france17} found a strong correlation between Ly$\alpha$ and the \h2 bump luminosity. However, Ly$\alpha$ cannot be observed directly and had to be reconstructed from other \h2 lines.  

Correlations have been seen between other FUV lines and accretion luminosity, L$_{acc}$, suggesting these lines are powered by the accretion process \citep[][RE19]{johns-krull00,calvet04,ingleby11a,yang12,gomezdecastro12, robinson19}.  CTTS have typical dipole field strengths of 0.5--1 kG \citep[e.g.,][]{donati09,johnskrull13} that are thought to be strong enough to truncate the inner disk and lead to the accretion of material onto the star via stellar magnetic field lines \citep{uchida85,koenigl91,shu94, hartmann16}. The funnel flow and accretion shock on the stellar surface produce near-ultraviolet (NUV), optical, and near-IR (NIR) continuum and line emission along with some X-ray emission. The most direct measurement of L$_{acc}$ (from which we measure the accretion rate, \mdot) comes from extracting the excess continuum emission from the accretion shock above the stellar photosphere. This excess is measured best in the NUV since there is less contribution there from the star \citep{ingleby11b}. The excess NUV and optical emission above the stellar photosphere has been fit with accretion shock models \citep{calvet98, herczeg08, rigliaco11, ingleby13, manara14}. Most of the X-ray photons emitted by the shock are expected to be absorbed.  However, {\it Chandra} and {\it XMM-Newton} observations detect an additional soft (0.5--1.5~keV) X-ray component (T~$\sim10^6$~K) that is much cooler than the coronal gas emission (T$\,\sim10^7$ K) in a few CTTS; this soft X-ray emission has been attributed to the accretion shock \citep[e.g.,][]{kastner02, stelzer04, schmitt05, gunther06, argiroffi07}.
 
Here we aim to search for correlations between L$_{acc}$, \mdot, the X-ray luminosity (L$_{X}$), and the \h2 bump luminosity in a sample of seven CTTS using multiple epochs of mostly simultaneous data from the {\it Hubble Space Telescope} ({\it HST}) and the {\it Neil Gehrels Swift Observatory} or the {\it Chandra X-ray Observatory}. \mdot\ (and hence L$_{acc}$) is known to vary \citep[e.g.,][RE19]{cody14,venuti14,ingleby15,cody18,siwak18}. X-ray emission from TTS is also known to be quite variable \citep[e.g.,][]{preibisch05,argiroffi11,flaccomio12,principe14,guarcello17}.  
However, the variability of the \h2 bump luminosity and its connection to the variability of both L$_{acc}$ and L$_{X}$ has not been explored previously, and this may help to understand the origin of the \h2 bump.  
We also test if there is any correlation in our sample between L$_{X}$ and accretion properties.

In Section 2, we present the data for our sample and provide a detailed overview of their simultaneity. In Section 3, we search for correlations between L$_{acc}$, \mdot, L$_{X}$, and the \h2 bump luminosity. In Section 4, we discuss the implications of the correlations we find and those we do not see. 

\begin{deluxetable*}{l l c c c c}[t]
\tabletypesize{\scriptsize}
\tablewidth{0pt}
\tablecaption{Log of $HST$ Observations}
\startdata
\hline
\colhead{Object} &\colhead{Epoch} & \colhead{Proposal ID} & \colhead{Date of Obs.} & \colhead{Start Time (UT)} & \colhead{End Time (UT)}\\
\hline
\hline
CS Cha & E1  & 13775  & 2015-04-23 & 01:07:08 & 03:34:10 \\
\hline 
DM Tau & E1  & 11608  & 2011-09-08 & 02:24:15 & 04:48:05 \\
DM Tau & E2  & 11608  & 2011-09-15 & 21:20:30 & 23:44:09 \\
DM Tau & E3  & 11608  & 2012-01-04 & 11:28:04 & 13:48:51 \\
\hline 
GM Aur & E1  & 11608 & 2011-09-11 & 18:17:51 & 20:39:14 \\
GM Aur & E2  & 11608 & 2011-09-17 & 21:19:50 & 23:43:00 \\
GM Aur & E3  & 11608 & 2012-01-05 & 06:40:14 & 09:01:56 \\
GM Aur & E4  & 14048 & 2016-01-05 & 20:38:03 & 23:00:42 \\
GM Aur & E5  & 14048 & 2016-01-09 & 13:40:55 & 16:02:33 \\
GM Aur & E6  & 15165 & 2018-01-04 & 06:10:54 & 08:34:18 \\
GM Aur & E7  & 15165 & 2018-01-11 & 05:03:19 & 07:26:40 \\
GM Aur & E8  & 15165 & 2018-01-19 & 03:43:47 & 06:07:09\\ 
\hline
SZ Cha & E1  & 13775 & 2015-03-15 & 02:18:17 & 04:36:17 \\
\hline 
Sz~45  & E1  & 14193 & 2016-05-14 & 20:11:30 & 22:35:27 \\
Sz~45  & E2  & 14193 & 2016-05-17 & 02:10:09 & 04:28:48 \\
Sz~45  & E3  & 14193 & 2016-05-18 & 17:50:27 & 19:58:51 \\
Sz~45  & E4  & 14193 & 2016-05-20 & 22:10:50 & \phn00:27:11$^{a}$ \\
Sz~45  & E5  & 14193 & 2016-07-06 & 01:06:12 & 03:33:04 \\
\hline 
TW Hya & E1  & 11608 & 2010-01-28 & 23:11:40 & \phn02:38:23$^{a}$ \\
TW Hya & E2  & 11608 & 2010-02-04 & 01:52:28 & 04:07:51 \\
TW Hya & E3  & 11608 & 2010-05-28 & 12:27:38 & 14:49:37 \\
TW Hya & E4  & 13775 & 2015-04-18 & 03:39:22 & 06:01:59 \\
\hline
VW Cha & E1  & 14193 & 2016-01-23 & 05:18:12 & 07:47:10 \\
VW Cha & E2  & 14193 & 2016-01-25 & 04:56:07 & 07:25:34 \\
VW Cha & E3  & 14193 & 2016-01-27 & 02:58:11 & 05:28:12 \\
VW Cha & E4  & 14193 & 2016-01-29 & 02:38:07 & 05:05:27 \\
VW Cha & E5  & 14193 & 2016-03-11 & 04:44:55 & 07:13:32 
\enddata
\tablenotetext{a}{Observations for Sz~45 E4 and TW Hya E1 ended on the subsequent UT dates (i.e., 2016-05-21 and 2010-01-29, respectively). } 
\label{tab:loghst}
\end{deluxetable*}

\begin{deluxetable*}{l l c c c c c}[t]
\tabletypesize{\scriptsize}
\tablewidth{0pt}
\tablecaption{Log of $Swift$ and $Chandra$ Observations}
\startdata
\hline
\colhead{Object} & \colhead{Epoch} & \colhead{Telescope} & \colhead{Obs.\ ID} & \colhead{Date of Obs.} & \colhead{Start Time (UT)} & \colhead{End Time (UT)}\\
\hline
\hline
CS Cha & E1  & {\it Swift} & 00032003002 & 2015-04-23 & 01:01:00 & 20:45:00 \\
\hline 
GM Aur & E4  & {\it Swift} & 00034249002 & 2016-01-05 & 20:43:02 & 22:58:00 \\
GM Aur & E4  & {\it Swift} & 00034249003 & 2016-01-06 & 00:00:00 & 14:42:00 \\
GM Aur & E5  & {\it Swift} & 00034249004 & 2016-01-09 & 09:13:26 & 16:13:53 \\
GM Aur & E6  & {\it Chandra} & 20614     & 2018-01-04 & 05:49:29 & 09:38:50    \\
GM Aur & E7  & {\it Chandra} & 20615     & 2018-01-11 & 04:45:07 & 08:32:11    \\
GM Aur & E8  & {\it Chandra} & 20616     & 2018-01-19 & 03:18:12 & 07:03:10    \\
\hline
SZ Cha & E1  & {\it Swift}  & 00033666001 & 2015-03-14 & 02:39:00 & 19:03:00  \\
SZ Cha & E1  & {\it Swift}  & 00033666002 & 2015-03-15 & 01:01:00 & 17:16:00 \\
\hline 
Sz~45 & E1  & {\it Swift}  & 00034501001 & 2016-05-14 & 17:24:00 & 00:00:00 \\
Sz~45 & E2  & {\it Swift}  & 00034501002 & 2016-05-17 & 07:21:00 & 13:58:00 \\
Sz~45 & E3  & {\it Swift}  & 00034501003 & 2016-05-18 & 16:54:00 & 21:51:00 \\
Sz~45 & E4  & {\it Swift}  & 00034501004 & 2016-05-20 & 16:53:00 & 23:29:00 \\
Sz~45 & E5  & {\it Swift}  & 00034501005 & 2016-07-06 & 03:40:00 & 14:57:00 \\
\hline 
TW Hya & E4  & {\it Swift}  & 00033736001 & 2015-04-18 & 04:29:00 & 04:53:00 \\
\hline
VW Cha & E1  & {\it Swift}  & 00034264001 & 2016-01-22 & 02:08:00 & \phn05:29:00$^{a}$ \\
VW Cha & E2  & {\it Swift}  & 00034283001 & 2016-01-25 & 01:48:02 & 09:50:22 \\
VW Cha & E3  & {\it Swift}  & 00034283002 & 2016-01-27 & 01:26:02 & 07:52:38 \\
VW Cha & E4  & {\it Swift}  & 00034283003 & 2016-01-29 & 01:07:36 & 05:57:39 \\
VW Cha & E5  & {\it Swift}  & 00034283004 & 2016-03-11 & 02:00:00 & 13:24:00
\enddata
\tablenotetext{a}{Observations for VW Cha E1 ended on the subsequent UT date (i.e., 2016-01-23).}
\label{tab:logxray}
\end{deluxetable*}

\section{Observations and Data Reduction} \label{redux} 

The goal of our study is to investigate how high-energy radiation fields affect gas in the innermost disk.  Most of our sample consists of objects previously identified as transitional or pre-transitional disks \citep[i.e., objects with large holes or gaps in the dust in their inner region; e.g.,][]{espaillat14}, and it has been seen that the \h2 bump is more often detected in transitional disks than full disks \citep{france17}. We note that VW~Cha is the only full disk in our sample, and it was included for comparison.  The objects in our sample also have {\it HST} data and/or X-ray observations from {\it Swift} or {\it Chandra} that were coordinated with {\it HST} observations. This results in a sample of seven TTS.  All targets have {\it HST} data.  Five targets have more than one epoch of {\it HST} data.  Six targets have coordinated X-ray and {\it HST} data.  Only DM Tau does not have coordinated X-ray data, but it is included in the sample because it has multiple epochs of {\it HST} data. To the best of our knowledge, these are all of the objects that have multiple epochs of {\it HST} FUV to NIR data or that have {\it HST} FUV to NIR data coordinated with X-ray observations that are currently available in the archive.

The start and end times of each of the {\it HST}, {\it Chandra}, and {\it Swift} observations are given in Tables \ref{tab:loghst} and \ref{tab:logxray} and are listed in order of the start time. Moving forward, we refer to observations with their object name and epoch (E), as listed in those tables. 

\subsection{{\it HST}} 

Our sample was observed with the Space Telescope Imaging Spectrograph (STIS) onboard {\it HST} (Table~\ref{tab:loghst}).  Spectra were obtained from the FUV to the NIR wavelengths (1100~{\AA}--1~{\micron}) using the MAMA detector with the G140L (1150~{\AA}--1730~{\AA}) and G230L (1570~{\AA}--3180~{\AA}) gratings and the CCD detector with the G430L (2900~{\AA}--5700~{\AA}) and G750L (5240~{\AA}--10,270~{\AA}) gratings. The spectra were obtained with a 52{\arcsec}$\times$2{\arcsec} slit, leading to resolutions (R) of $\sim500$--1440 for the G140L and G230L gratings and R$\,\sim530$--1040 for the G430L and G750L gratings. The only exception to the above is TW~Hya, which is too bright in the FUV for the G140L grating.  In the case of TW~Hya, the E140M (1144~{\AA}--1710~{\AA}) grating was used with a 0.2\arcsec$\times0.2$\arcsec slit, for a resolution of about 45,800.  Here we convolve the TW~Hya E140M spectra to match the resolution of the G140L data to facilitate comparison.

The {\it HST} data for CS~Cha and SZ~Cha are presented here for the first time.  The {\it HST} data for the other objects in our sample were presented in RE19.  We refer the reader to that paper for further details on the exposure times and data reduction. We note that GM Aur E1, E2, and E3 were also presented previously in \citet{ingleby15}.  

For CS~Cha, exposure times with the G140L, G230L, G430L, and G750L gratings were 3315~s, 1178~s, 303~s, and 328~s, respectively. For SZ~Cha, exposure times with the gratings were 3315~s, 1491~s, 20~s, and 2~s, respectively. Data were obtained from the STScI {\it calstis} reduction pipeline.  We corrected the G750L spectra for fringing that typically occurs at wavelengths longer than approximately 7000~{\AA} by following the steps outlined in \citet{goudfrooij98} using a contemporaneous flat that was taken alongside the science observations. After the fringes were removed, the product was passed through the standard {\it HST} STIS pipeline to complete calibration. 

\begin{deluxetable*}{l l cc c c c c}[t]
\tabletypesize{\scriptsize}
\tablewidth{0pt}
\tablecaption{X-ray Spectral Fitting Results}
\startdata
\hline
\colhead{Object} & \colhead{Epoch} & \colhead{Exp. Time } & \colhead{Net Count Rate} & \colhead{C-Statistic} & \colhead{Degrees of} &\colhead{$kT$} & \colhead{Unabsorbed Flux} \\
\colhead{} & \colhead{} & \colhead{(s)} & \colhead{(10$^{-2}$ cts$/$s)} & \colhead{or $\chi^2$ Value$^{a}$} & \colhead{Freedom} & \colhead{(keV)} & \colhead{(10$^{-12}$ erg s$^{-1}$ cm$^{-2}$)} \\
\hline

CS Cha & E1 & 7052 & 3.47$_{-0.22}^{+0.22}$ & 180.56\phn & 129\phn & 1.00$_{-0.06}^{+0.05}$ & 0.87$_{-0.09}^{+0.10}$\\
\hline

GM Aur & E4 & 4900 & 2.1$_{-0.2}^{+0.2}$& 87.73 & 42 & 0.45$_{-0.12}^{+0.14}$ & 1.11$_{-0.19}^{+0.29}$\\
GM Aur & E5 & 11910\phn & 9.7$_{-0.3}^{+0.3}$ & 422.92\phn & 394\phn & 4.2$_{-0.5}^{+0.6}$ & 5.6$_{-0.3}^{+0.4}$\\

GM Aur & E6$^{b}$ & 10530\phn  & 3.6$_{-0.2}^{+0.2}$ & -- & -- & -- & 1.45$_{-0.16}^{+0.14}$ \\
GM Aur & E7$^{b}$ & 10490\phn  & 2.9$_{-0.2}^{+0.2}$ & 88.5 & 78 & 0.17$_{-0.07}^{+0.04}$, 0.94$_{-0.08}^{+0.08}$& 1.35$_{-0.18}^{+0.14}$ \\

GM Aur & E8$^{b}$ & 10520\phn  & 3.5$_{-0.2}^{+0.2}$ & -- & -- & -- & 1.55$_{-0.19}^{+0.15}$ \\
\hline

SZ Cha & E1 & 23180\phn & 0.87$_{-0.06}^{+0.06}$ & 385.3\phn & 752\phn & 2.0$_{-0.4}^{+0.5}$ & 0.26$_{-0.04}^{+0.04}$\\
\hline

Sz~45 & E1 & 3279 & 0.63$_{-0.14}^{+0.14}$ & 31.21 & 18 & 2.7$_{-1.2}^{+7.0}$ & 0.24$_{-0.08}^{+0.10}$\\
Sz~45 & E2 & 6008 & 0.79$_{-0.12}^{+0.12}$ & 62.36 & 42 & 0.83$_{-0.05}^{+0.18}$ & 0.19$_{-0.05}^{+0.06}$\\
Sz~45 & E3 & 5676 & 0.92$_{-0.13}^{+0.13}$ & 62.20 & 41 & 1.00$_{-0.22}^{+0.29}$ & 0.22$_{-0.05}^{+0.06}$\\
Sz~45 & E4 & 5983 & 0.63$_{-0.11}^{+0.11}$ & 31.79 & 33 & \phn5$_{-4}^{+60}$ & 0.36$_{-0.17}^{+0.23}$\\
Sz~45 & E5 & 7834 & 0.73$_{-0.10}^{+0.10}$ & 65.22 & 48 & 1.9$_{-0.9}^{+1.0}$ & 0.22$_{-0.06}^{+0.07}$\\
\hline

TW Hya & E4 & 1321 & 20.7$_{-1.3}^{+1.3}$ & 94.60 & 11 & 0.77$_{-0.06}^{+0.06}$ & 5.0$_{-0.5}^{+0.5}$\\
\hline

VW Cha & E1 & 7719 & 2.78$_{-0.19}^{+0.19}$ & 117.17\phn & 130\phn & 2.04$_{-0.29}^{+0.60}$ & 1.00$_{-0.13}^{+0.18}$\\
VW Cha & E2 & 3744 & 4.0$_{-0.3}^{+0.3}$ & 118.30\phn & 118\phn & 6.3$_{-2.4}^{+6.0}$ & 2.1$_{-0.4}^{+0.4}$\\
VW Cha & E3 & 4001 & 2.39$_{-0.25}^{+0.25}$ & 69.59 & 78 & 2.2$_{-0.5}^{+0.9}$ & 0.82$_{-0.17}^{+0.19}$\\
VW Cha & E4 & 3632 & 7.0$_{-0.4}^{+0.4}$ & 145.30\phn & 193\phn & 60$_{-40}^{+60}$ & 4.4$_{-0.6}^{+0.5}$\\
VW Cha & E5 & 9385 & 2.59$_{-0.17}^{+0.17}$ & 158.61\phn & 156\phn &  3.3$_{-0.7}^{+1.0}$ & 1.00$_{-0.14}^{+0.16}$
\enddata
\tablecomments{
$N_H$ was adopted from \citet{kalberla05} and can be found in Section~\ref{redux:swift}. 
}
\tablenotetext{a}{For the {\it Swift} observations, we list the C-statistic.  For the {\it Chandra} observations, we list the $\chi^2$ value.}
\tablenotetext{b}{We fit all the {\it Chandra} spectra with the same two-temperature model.  We list the $\chi^2$ value, degrees of freedom, and temperatures for the joint fit to all three spectra.}
\label{tab:xrayflux}
\end{deluxetable*}

\subsection{{\it Swift}} \label{redux:swift}

{\it Swift} observations of CS~Cha, GM~Aur, SZ~Cha, Sz~45, TW~Hya, and VW~Cha were taken with the X-ray Telescope on the dates listed in Table~\ref{tab:logxray}. We utilized the High Energy Astrophysics Science Archive Research Center (HEASARC) HEASOFT software (v.~6.22.1) to analyze the {\it Swift} data. We fit our data with the X-ray spectral fitting package XSPEC \citep{arnaud96} using one Astrophysical Plasma Emission Code (APEC) model. Values for neutral hydrogen columns ($N_H$) for each of our targets were obtained from the Leiden/Argentine/Bonn survey \citep{kalberla05}. We adopted an $N_H$ for CS~Cha, GM~Aur, SZ~Cha, Sz~45, TW~Hya, and VW~Cha of $7.74\times10^{20}$~cm$^{-2}$, $2.51\times10^{21}$~cm$^{-2}$, $7.68\times10^{20}$~cm$^{-2}$, $7.45\times10^{20}$~cm$^{-2}$, $5.43\times10^{20}$~cm$^{-2}$, and $7.83\times10^{20}$~cm$^{-2}$, respectively, as the input for the modifying absorption component, \textit{phabs}. 
We used the C-statistic to judge a goodness of fit for the model. We present the exposure times, net count rates, C-statistic, degrees of freedom of the fit, $kT$, and the unabsorbed X-ray fluxes in Table~\ref{tab:xrayflux}. Uncertainties are reported at the 90$\%$ confidence level.

Values reported in Table~\ref{tab:xrayflux} for GM~Aur E4 and SZ~Cha E1 were calculated by combining data from multiple observations (Table~\ref{tab:logxray}). In the case of GM~Aur, the data from Obs.\ ID 00034249002 overlap with the {\it HST} observations, and the data from Obs.\ ID 00034249003 were taken significantly later  (see Section~\ref{sec:simulaneity}).  However, the flux obtained from combining the two observations is similar to the individual fluxes. Using {\it addascaspec} to add the data from Obs.\ ID 00034249002 and Obs.\ ID 00034249003, we measure a combined X-ray flux of about $1.11_{-0.19}^{+0.29}\times10^{-12}$ erg s$^{-1}$ cm$^{-2}$ (Table~\ref{tab:xrayflux}), which is similar to the individual fluxes obtained for Obs.\ ID 00034249002 and Obs.\ ID 00034249003 of 
$1.2_{-0.2}^{+0.2}\times10^{-12}$ erg s$^{-1}$ cm$^{-2}$ and  
$0.62_{-0.06}^{+0.06}\times10^{-12}$ erg s$^{-1}$ cm$^{-2}$, respectively. 
Therefore, we use the flux from the combined observations for GM~Aur E4 moving forward.
Similarly, the SZ~Cha data from Obs.\ ID 00033666002 were taken simultaneously with the {\it HST} observations, but the data from Obs.\ ID 00033666001 were taken earlier.  We find that the combined X-ray flux of the two observations is 
$0.26_{-0.04}^{+0.04}\times10^{-12}$ erg s$^{-1}$ cm$^{-2}$ (Table~\ref{tab:xrayflux}), which is similar to the individual fluxes for Obs.\ ID 00033666001 and Obs.\ ID 00033666002 of 
$0.21_{-0.03}^{+0.03}\times10^{-12}$ erg s$^{-1}$ cm$^{-2}$ and 
$0.38_{-0.04}^{+0.05}\times10^{-12}$ erg s$^{-1}$ cm$^{-2}$, respectively; we thus use the combined flux for SZ~Cha E1 moving forward.

\subsection{{\it Chandra}} \label{redux:chandra}

{\it Chandra} observations were obtained as part of a joint {\it HST--Chandra} GO program for GM~Aur ({\it HST} Observation ID 15165, {\it Chandra} Observation IDs 20614, 20615, 20616). The \textit{Chandra} observations were performed with the Advanced CCD Imaging Spectrometer (ACIS). The dates and times of the observations can be found in Table~\ref{tab:logxray}.   GM Aur was placed on the ACIS-S3 detector at the nominal aimpoint. The detector was operated in VFAINT mode with a 1/8 subarray option in order to mitigate potential pile-up due to the known variable X-ray emission from the target (t$_{readout}$ = 0.441~s). Spectra and light curves were extracted from the standard pipeline processed level II event files (\textsc{ASCDSVER} = 10.6). Spectra were extracted using the \textit{specextract} task in \textsc{CIAO 4.8} with \textsc{CALDB 4.7.4} and were grouped to have a S/N of 3 per bin. Source counts were extracted from a 5{\arcsec} radius circular region centered on the known position of GM~Aur.  Background counts were extracted from an annular region centered on the source  position with inner and outer radii of 7.4{\arcsec} and 14.8{\arcsec}, respectively. 

\begin{figure}
\epsscale{1.0}
\plotone{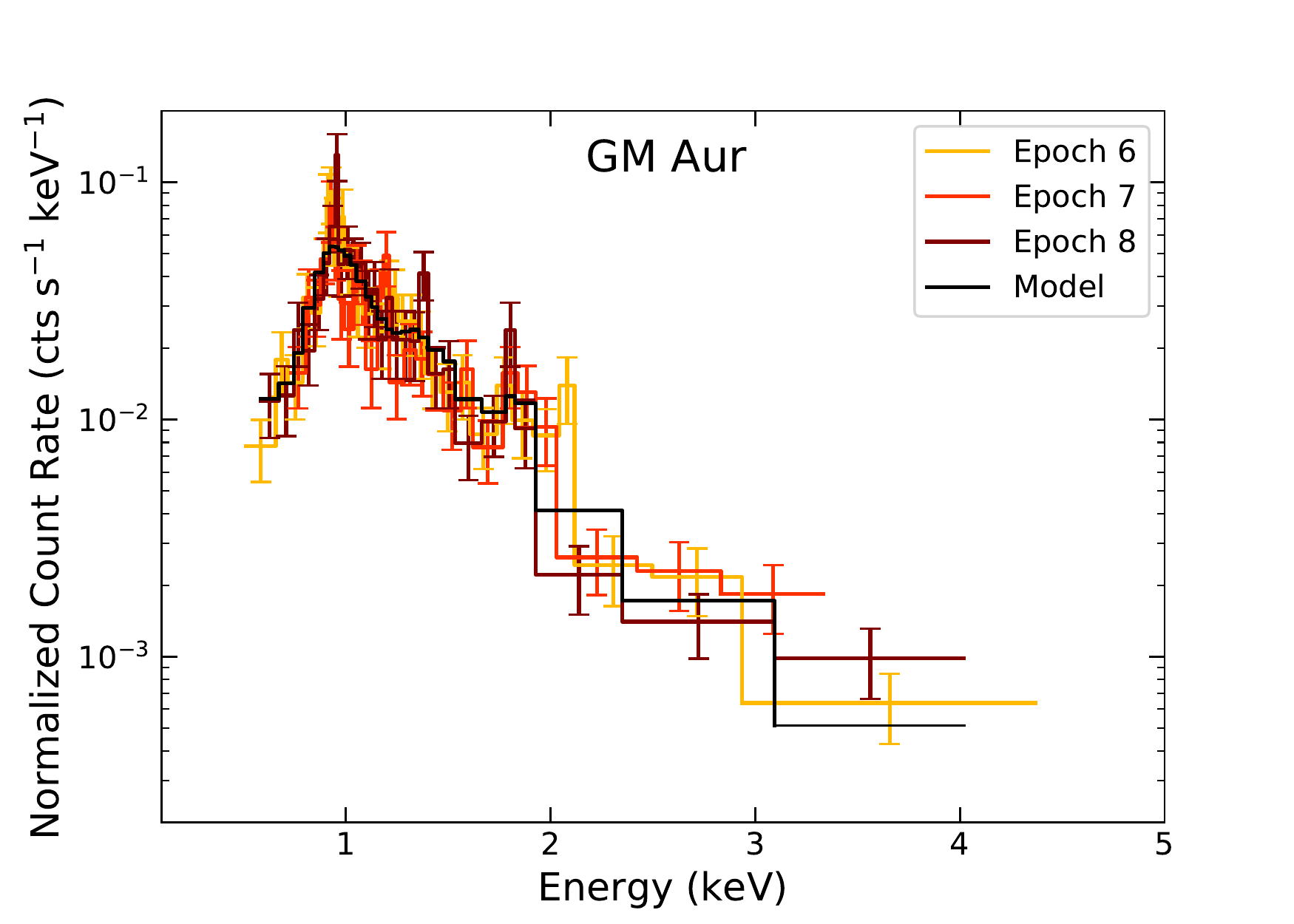}
\caption[]{
{\it Chandra} ACIS-S3 spectra of GM~Aur E6, E7, and E8. We fit all epochs with the same two-temperature APEC model (black line).  Best-fitting model parameters are presented in Table~$\ref{tab:xrayflux}$. 
}
\label{fig:xspec}
\end{figure}

X-ray light curves were created with \textit{dmsextract} using the previous extraction regions. We fit the spectra in XSPEC (v.~12.9.1) with two-temperature APEC thermal collisional ionization equilibrium plasma models \citep{smith01} along with an absorbing column of interstellar material  \citep[i.e., {\it tbabs} absorption model;][]{wilms00}. We used the same absorbing hydrogen column density as we did for the {\it Swift} reduction of GM~Aur (Section~\ref{redux:swift}). The fits assume the same temperature for each epoch for the soft and hard components, with the normalization allowed to vary. The abundance is allowed to vary but is tied between all components; we find $z=0.19_{-0.07}^{+0.09}$. X-ray light curves (Figure~\ref{fig:xspec}) are discussed further in Section~\ref{sec:corr}. 
In Table~\ref{tab:xrayflux}, we list exposure times, net count rates, $\chi^2$ values, degrees of freedom of the fit, $kT$, and unabsorbed X-ray fluxes. We note that uncertainties are at the 90\% confidence level. 

\subsection{Simultaneity of the Observations} \label{sec:simulaneity}

For this study, we use data from {\it HST}, {\it Chandra}, and {\it Swift} to meaure the \h2 bump luminosity, L$_{acc}$, and L$_X$. The H$_2$ luminosity is measured from {\it HST} data in this work; L$_{acc}$ is calculated using stellar parameters taken from the literature, and \mdot\ is either adopted from RE19 or measured in the Appendix; L$_X$ is measured in this work from either {\it Swift} or {\it Chandra} data. Here we review the timing of the {\it HST} and X-ray observations in order to assess how much of these data were taken simultaneously (i.e., observed at the same time) as opposed to contemporaneously (i.e., typically taken to mean within a day or so in the literature). 

The \h2 bump is located in the FUV while L$_{acc}$ is dominated by NUV emission.  These are covered by separate gratings that were taken in adjacent {\it HST} orbits. The typical length of {\it HST} observations is 2.5~hrs over two orbits. This is important to note since variability on short timescales of a few minutes has been observed in CTTS and has been associated with accretion \citep[e.g.,][]{cody14, siwak18}.  However, RE19 find that for the {\it HST} spectra used in this work, the flux agrees in the wavelength regions that overlap at the edges of the gratings that were taken in different orbits, suggesting minimal discernible variability within the individual sets of observations.  Therefore, moving forward, we assume there was no significant variability within the 2.5~hrs of the {\it HST} observations.

In general, when {\it Chandra} data are available, they were taken simultaneously with {\it HST} over the length of the {\it HST} observations. Most {\it HST} and {\it Swift} data were simultaneous for some portion.  However, given the Target of Opportunity nature of the {\it Swift} science program, it was not possible to guarantee strictly simultaneous, uninterrupted observations.  

We discuss the timing of the observations in detail for each of the objects below.  In sum, the majority of the {\it HST} and X-ray data sets used in our analysis are at least partially simultaneous.  Of the 18 epochs of coordinated {\it HST} and X-ray data here, 4 were entirely simultaneous (GM~Aur E6, E7, E8, TW~Hya E4), 8 are partially simultaneous with the rest of the data taken within 6 hrs (GM~Aur E5, Sz~45 E1, E3, E4, VW~Cha E2, E3, E4, E5), and 3 are partially simultaneous with the rest of the data taken within 6--21 hrs (CS~Cha E1, GM~Aur E4, VW~Cha E1).  Only three data sets did not overlap at all with the {\it HST} observations:  Sz~45 E2 and E5 were taken within 12~hrs of the {\it HST} observations;  and SZ~Cha E1 was taken within 24~hrs of the {\it HST} observations.  Moving forward, we refer to our overall sample as mostly simultaneous.

\subsubsection{CS Cha}

We have one epoch of {\it Swift} observations for CS~Cha.  About 20\% of this observation was simultaneous with the {\it HST} observations.  The rest of the {\it Swift} data were taken within 17~hrs after the end of the {\it HST} observations. 

\subsubsection{DM Tau}

DM~Tau does not have coordinated {\it HST} and X-ray observations.

\subsubsection{GM Aur}

GM~Aur E1, E2, and E3 do not have coordinated {\it HST} and X-ray observations.
For GM~Aur E4 of the {\it HST} observations, the {\it Swift} data from 2016 Jan~5 (Obs.\ ID 00034249002) were taken entirely within the {\it HST} observation time.  The {\it Swift} data from 2016 Jan~6 (Obs.\ ID 00034249003) were taken within 16~hrs after the {\it HST} observations from E4 ended.  In Section~\ref{redux:swift}, we show that the fluxes from these Obs.\ IDs were very similar, and so we use the combined flux in this work.  For GM~Aur E5, about 40\% of the {\it Swift} observations were simultaneous with {\it HST}. The rest of the {\it Swift} data in E5 were taken less than 4.5~hrs before the start of the {\it HST} observations.

We have {\it Chandra} data for GM~Aur E6, E7, and E8.  Our {\it HST} and {\it Chandra} data were taken simultaneously over the length of the {\it HST} observations. The {\it Chandra} data generally began 20--25 minutes before the {\it HST} observations and ended about one hour after the {\it HST} observations. GM~Aur is the only target that has {\it Chandra} observations.

\subsubsection{SZ Cha}

We have two {\it Swift} observations for SZ~Cha. Data from Obs.\ ID 00033666001 were taken within 24 hrs before the start of the {\it HST} observations.  About 25\% of the Obs.\ ID 00033666002 observations were simultaneous with the {\it HST} observations.  The rest of the {\it Swift} data were taken either 1~hr before the start of or within 13~hrs after the end of the {\it HST} observations.

\subsubsection{Sz 45}

We have five {\it Swift} observations for Sz~45.
For E1, about 40\% of these observations were simultaneous with the {\it HST} observations.  The rest of the {\it Swift} data were taken within 3~hrs before the start of and 2~hrs after the end of the {\it HST} observations. 
For E2, none of these observations were simultaneous with the {\it HST} observations.  The {\it Swift} data were taken within 10 hrs after the end of the {\it HST} observations. 
For E3, about 30\% of these observations were simultaneous with the {\it HST} observations.  The rest of the {\it Swift} data were taken within 1~hr before the start of and 2~hrs after the end of the {\it HST} observations. 
For E4, about 10\% of these observations were simultaneous with the {\it HST} observations.  The rest of the {\it Swift} data were taken within 6~hrs before the start of the {\it HST} observations. 
For E5, none of these observations were simultaneous with the {\it HST} observations.  The {\it Swift} data were taken within 12~hrs after the end of the {\it HST} observations.

\subsubsection{TW Hya}

We do not have coordinated {\it HST} and X-ray observations for E1, E2, and E3 of TW~Hya.
For TW~Hya E4, the entirety of the {\it Swift} observations were simultaneous with the {\it HST} observations.  

\begin{deluxetable*}{l l c c c c c c }[t]
\tabletypesize{\scriptsize}
\tablewidth{0pt}
\tablecaption{Adopted Stellar Parameters}
\startdata
\hline
\colhead{Object} & \colhead{Distance} & \colhead{A$_V$} & \colhead{SpT} & \colhead{T$_{*}$} & \colhead{M$_{*}$} & \colhead{R$_{*}$} & \colhead{L$_{*}$} \\
\colhead{} & \colhead{(pc)} & \colhead{} & \colhead{} & \colhead{(K)} & \colhead{($\msun$)} & \colhead{($\rsun$)} & \colhead{($\lsun$)}\\
\hline
\hline
CS Cha 	& 176.3$\pm$1.2 	& 0.8 & K2	& 4900 	& 1.32 & 1.83 & 1.75\\
DM Tau 	& 145.1$\pm$1.1 	& 1.1 & M2	& 3560	& 0.56 & 1.63 & 0.39 \\
GM Aur 	& 159.6$\pm$2.1 	& 0.6 & K5 	& 4350	& 1.36 & 2.0\phn & 1.29 \\
SZ Cha 	& 189.8$\pm$1.5	& 1.3 & K2	& 4900	& 1.22 & 1.78 & 1.66 \\
Sz~45 	& 188.4$\pm$0.9 	& 0.7 & M0.5	& 3780 	& 0.85 & 1.78 & 0.59 \\ 
TW Hya 	& 60.09$\pm$0.15 	& 0.0 & K7 	& 4060	& 0.79 & 0.93 & 0.21 \\
VW Cha 	& 190$\pm$5	& 1.9 & K7 	& 4060	& 1.24 & 3.08\phn & 2.34
\enddata
\tablecomments{Distances are from \textit{Gaia} DR2 \citep{gaia18b} except for VW~Cha, for which we adopt the median distance to Cha~I measured from \textit{Gaia} data \citep{roccatagliata18}.  Other stellar parameters are adopted from \citet{manara14} except for VW~Cha \citep{manara17}; both works use the \citet{baraffe98} stellar evolution tracks.  We note that radii and luminosities have been scaled to reflect the \textit{Gaia} distances. Temperatures are adopted from \citet{kh95} for the corresponding spectral type.}
\label{tab:starparam}
\end{deluxetable*}

\subsubsection{VW Cha}

We have five {\it Swift} observations for VW~Cha.  
For E1, about 10\% of these observations were simultaneous with the {\it HST} observations. The rest of the {\it Swift} data were taken within about  27~hrs before the start of the {\it HST} observations.
For E2, 40\% of these observations were simultaneous with the {\it HST} observations.  The {\it Swift} data were taken within 3 hrs before the start of and 3~hrs after the end of the {\it HST} observations.
For E3, about 45\% of these observations were simultaneous with the {\it HST} observations.  The rest of the {\it Swift} data were taken within 2~hrs before the start of and 3~hrs after the end of the {\it HST} observations.
For E4, about 70\% of these observations were simultaneous with the {\it HST} observations.  The rest of the {\it Swift} data were taken within 2~hrs before the start of and 1~hr after the end of the {\it HST} observations.
For E5, 30\% of these observations were simultaneous with the {\it HST} observations.  The {\it Swift} data were taken within 3~hrs before the start of and 6~hrs after the end of the {\it HST} observations.

\section{Analysis \& Results} \label{sec:analysis}

Here we present the adopted stellar parameters of our sample (i.e., extinction, distance, stellar luminosity, spectral type, accretion rate).  For each epoch  of observations, we also provide measurements of L$_X$, L$_{acc}$, and the luminosity of the \h2 bump as well as two \h2 emission lines. These properties are measured using {\it Swift}/{\it Chandra}, {\it HST} NUV, and {\it HST} FUV data, respectively. Finally, we search for correlations between the above-mentioned properties, including comparisons between L$_X$ and UV line luminosities and accretion column properties previously reported by RE19.

\begin{deluxetable*}{l l c c c c c c}[t]
\tabletypesize{\scriptsize}
\tablewidth{0pt}
\tablecaption{Measured Gas and X-ray Properties}
\startdata
\hline
\colhead{Object} & \colhead{Epoch} & \colhead{\mdot} & \colhead{L$_{acc}$} & \colhead{L$_{X}$} &  \colhead{L (\h2 bump)} &  \colhead{L (1-7R(3))} &  \colhead{L (B-X(5-12)P(3))} \\
\colhead{} & \colhead{} & \colhead{(10$^{-8}$$\msunyr$)} & \colhead{($\lsun$)} & \colhead{($10^{30} {\rm erg s}^{-1}$)} & \colhead{($10^{29} {\rm erg s}^{-1}$)} & \colhead{($10^{29} {\rm erg s}^{-1}$)} & \colhead{($10^{29} {\rm erg s}^{-1}$)}\\
\hline
\hline

CS Cha & E1 & $1.497^{+0.010}_{-0.009}$ &
$0.274^{+0.002}_{-0.002}$ & $3.2^{+0.4}_{-0.3}$ &33.6$^{+3.7}_{-3.9}$
& $6.7^{+1.7}_{-1.7}$ & $ 2.1^{+0.5}_{-0.5}$\\

\hline 
DM Tau & E1 & $2.770^{+0.026}_{-0.025}$ &
0.2421$^{+0.0023}_{-0.0022}$ & -- &45$^{+6}_{-6}$
& $ 3.6 ^{+ 0.9 }_{- 0.9 }$ & $ 0.8 ^{+ 0.2 }_{- 0.2 }$\\
DM Tau & E2 & $3.582^{+0.029}_{-0.029}$ &
0.3130$^{+0.0025}_{-0.0025}$ & -- &43$^{+7}_{-7}$
& $ 2.6 ^{+ 0.6 }_{- 0.6 }$ & $ 0.7 ^{+ 0.2 }_{- 0.2 }$\\
DM Tau & E3 & $2.011^{+0.028}_{-0.027}$ &
0.1758$^{+0.0025}_{-0.0024}$ & -- &40$^{+7}_{-7}$
& $ 4.3 ^{+ 1.1 }_{- 1.1 }$ & $ 1.6 ^{+ 0.4 }_{- 0.4 }$\\

\hline 
GM Aur & E1 & $1.546^{+0.011}_{-0.011}$ &
0.268$^{+0.002}_{-0.002}$ & -- & 11.1$^{+1.2}_{-1.2}$
& $ 0.9 ^{+ 0.2 }_{- 0.2 }$ & $ 0.3 ^{+ 0.1 }_{- 0.1 }$\\
GM Aur & E2 & $1.291^{+0.010}_{-0.010}$ &
0.224$^{+0.002}_{-0.002}$ & -- & 11.5$^{+1.1}_{-1.2}$
& $ 0.8 ^{+ 0.2 }_{- 0.2 }$ & $ 0.17 ^{+ 0.04 }_{- 0.04 }$\\
GM Aur & E3 & $0.660^{+0.007}_{-0.007}$ &
0.114$^{+0.001}_{-0.001}$ & -- &8.9$^{+0.9}_{-1.0}$
& $ 0.7 ^{+ 0.2 }_{- 0.2 }$ & $ 0.6 ^{+ 0.1 }_{- 0.1 }$\\
GM Aur & E4 & $1.021^{+0.009}_{-0.009}$ &
0.177$^{+0.002}_{-0.002}$ & $3.4^{+0.9}_{-0.6}$ & 9.4$^{+1.2}_{-1.2}$
& $ 1.0 ^{+ 0.3 }_{- 0.3 }$ & $ 0.5 ^{+ 0.1 }_{- 0.1 }$\\
GM Aur & E5 & $0.768^{+0.008}_{-0.008}$ &
0.133$^{+0.001}_{-0.001}$ & $17.1^{+1.2}_{-0.9}$ & 11.3$^{+1.3}_{-1.3}$
& $ 1.3 ^{+ 0.3 }_{- 0.3 }$ & $ 0.4 ^{+ 0.1 }_{- 0.1 }$\\
GM Aur & E6 & $0.564^{+0.007}_{-0.007}$ &
0.098$^{+0.001}_{-0.001}$ & $4.4^{+0.4}_{-0.5}$ & 9.1$^{+1.0}_{-1.0}$
& $ 1.1 ^{+ 0.3 }_{- 0.3 }$ & $ 0.4 ^{+ 0.1 }_{- 0.1 }$\\
GM Aur & E7 & $1.961^{+0.012}_{-0.012}$ &
0.339$^{+0.002}_{-0.002}$ & $4.1^{+0.4}_{-0.6}$ & 17.1$^{+2.0}_{-2.0}$
& $ 2.3 ^{+ 0.6}_{- 0.6 }$ & $ 0.13 ^{+ 0.03 }_{- 0.03 }$\\
GM Aur & E8 & $0.979^{+0.009}_{-0.009}$ &
0.170$^{+0.002}_{-0.002}$ & $4.7^{+0.5}_{-0.6}$ & 10.1$^{+1.3}_{-1.3}$
& $ 1.1 ^{+ 0.3 }_{- 0.3 }$ & $ 0.4 ^{+ 0.1 }_{- 0.1 }$\\

\hline
SZ Cha  & E1 & $0.354^{+0.011}_{-0.009}$ &
$0.062^{+0.002}_{-0.002}$ & $1.1^{+0.2}_{-0.2}$ & 4.5$^{+2.6}_{-2.6}$
& $ 0.9 ^{+ 0.2 }_{- 0.2 }$ & $ 0.08 ^{+ 0.02 }_{- 0.02 }$\\

\hline 
Sz 45 & E1 & $0.923^{+0.015}_{-0.015}$ &
0.112$^{+0.002}_{-0.002}$ & $1.0^{+0.4}_{-0.3}$  & 5.0$^{+1.6}_{-1.6}$
& $ 1.0 ^{+ 0.3 }_{- 0.3 }$ & $ 0.11 ^{+ 0.03 }_{- 0.03 }$\\
Sz 45 & E2 & $1.186^{+0.018}_{-0.017}$ &
0.144$^{+0.002}_{-0.002}$ & $0.8^{+0.3}_{-0.2}$  & 1.6$^{+1.4}_{-1.4}$
& $ 0.8 ^{+ 0.2 }_{- 0.2 }$ & $ 0.15 ^{+ 0.04 }_{- 0.04 }$\\
Sz 45 & E3 & $1.490^{+0.019}_{-0.019}$ &
0.181$^{+0.002}_{-0.002}$ & $1.0^{+0.3}_{-0.3}$  & 2.6$^{+0.6}_{-0.6}$
& $ 0.7 ^{+ 0.2 }_{- 0.2 }$ & $ 0.06 ^{+ 0.02 }_{- 0.02 }$\\
Sz 45 & E4 & $1.639^{+0.020}_{-0.020}$ &
0.199$^{+0.002}_{-0.002}$ & $1.5^{+1.0}_{-0.7}$  & 2.3$^{+1.0}_{-1.0}$
& $ 0.8 ^{+ 0.2 }_{- 0.2 }$ & $ 0.09 ^{+ 0.03 }_{- 0.03 }$\\
Sz 45 & E5 & $1.157^{+0.018}_{-0.017}$ &
0.140$^{+0.002}_{-0.002}$ & $0.9^{+0.3}_{-0.3}$ & 3.9$^{+1.0}_{-1.0}$
& $ 0.9 ^{+ 0.2 }_{- 0.2 }$ & $ 0.13 ^{+ 0.03 }_{- 0.03 }$\\
 
\hline 
TW Hya & E1 & $0.330^{+0.004}_{-0.004}$ &
0.071$^{+0.001}_{-0.001}$ & -- & 5.1$^{+1.3}_{-1.3}$
& $ 0.17 ^{+ 0.04 }_{- 0.04 }$ & $ 0.16 ^{+ 0.04 }_{- 0.04 }$\\
TW Hya & E2 & $0.1384^{+0.0028}_{-0.0030}$ &
0.0299$^{+0.0006}_{-0.0007}$ & --  & 3.6$^{+1.5}_{-1.5}$
& $ 0.28 ^{+ 0.07 }_{- 0.07 }$ & $ 0.18 ^{+ 0.05 }_{- 0.05 }$\\
TW Hya & E3 & $0.2206^{+0.0029}_{-0.0040}$ &
0.0477$^{+0.0006}_{-0.0008}$ & --  & 4.2$^{+1.4}_{-1.4}$
& $ 0.18 ^{+ 0.05 }_{- 0.05 }$ & $ 0.07 ^{+ 0.02 }_{- 0.02 }$\\
TW Hya & E4 & $0.262^{+0.005}_{-0.004}$ &
0.057$^{+0.001}_{-0.001}$ & $2.2^{+0.2}_{-0.2}$ &4.6$^{+1.3}_{-1.3}$
& $ 0.24 ^{+ 0.06 }_{- 0.06 }$ & $ 0.08 ^{+ 0.02 }_{- 0.02 }$\\

\hline
VW Cha & E1 & $8.5^{+0.3}_{-2.0}$ &
0.87$^{+0.03}_{-0.2}$ & $4.3^{+0.8}_{-0.6}$ & 45.7$^{+32.5}_{-35.4}$
& $ 8.4 ^{+ 2.1 }_{- 2.1 }$ & $ 5.21 ^{+ 1.3 }_{- 1.3 }$\\
VW Cha & E2 & $7.46^{+0.2}_{-0.2}$ &
0.76$^{+0.02}_{-0.02}$ & $9.1^{+1.7}_{-1.7}$ & 56.9$^{+24.1}_{-24.4}$
& $ 4.4 ^{+ 1.1 }_{- 1.1 }$ & $ 1.5 ^{+ 0.4 }_{- 0.4 }$\\
VW Cha & E3 & $15.1^{+0.3}_{-0.3}$ &
1.55$^{+0.03}_{-0.03}$ & $3.5^{+0.8}_{-0.7}$ & 85.5$^{+39.2}_{-38.5}$
& $ 6.2 ^{+ 1.6 }_{- 1.6 }$ & $ 1.4 ^{+ 0.3 }_{- 0.3 }$\\
VW Cha & E4 & $19.9^{+0.4}_{-0.4}$ &
2.03$^{+0.04}_{-0.04}$ & $19.0^{+2.2}_{-2.6}$ & 119.7$^{+46.5}_{-47.5}$
& $ 4.6 ^{+ 1.1 }_{- 1.1 }$ & $ 3.0 ^{+ 0.7 }_{- 0.7 }$\\
VW Cha & E5 & $6.91^{+0.06}_{-0.05}$ &
0.706$^{+0.007}_{-0.005}$ & $4.3^{+0.7}_{-0.6}$ & 74.2$^{+27.5}_{-27.5}$
& $ 5.0 ^{+ 1.3 }_{- 1.3 }$ & $ 1.2 ^{+ 0.3 }_{- 0.3 }$

\enddata
\tablecomments{All \h2 bump luminosities have been updated.  In addition, for VW Cha, all other parameters have been updated. We add two new columns for two H$_{2}$ line luminosity measurements. Accretion rates and luminosities are adopted from RE19 except for CS~Cha and SZ~Cha, which we measure in the Appendix. As noted by RE19, the uncertainties listed here do not take into account broader systematic uncertainties (e.g., extinction) and reflect only the width of the marginalized posterior from an MCMC analysis. RE19 note that the uncertainties in the \mdot\ are about 10\%, assuming a visual extinction correction error of 0.5. All X-ray, \h2 bump, \h2 1-7R(3), and \h2 B-X(5-12)P(3) luminosities are measured in this work.}
\label{tab:values}
\end{deluxetable*}

\subsection{Stellar Properties} \label{sec:star}

We follow RE19 and adopt distances, visual extinctions (A$_{V}$), spectral types, stellar temperatures, masses, radii, and luminosities from the same literature sources (Table~\ref{tab:starparam}).  We adopt \mdot\ (Table~\ref{tab:values}) from RE19 except for CS~Cha and SZ~Cha, whose accretion properties are derived in the Appendix following the methods of RE19.  
We calculate L$_{acc}$ using 
 \begin{equation}
L_{acc}=\frac{G M_* \mdot}{R_*} \left ( 1 - \frac{R_*}{R_{in}} \right )
 \end{equation}
with the values listed in Table~\ref{tab:values} and $R_{in}=5R_{*}$.
Using the X-ray fluxes from Table~\ref{tab:xrayflux} and distances (Table~\ref{tab:starparam}), we calculate X-ray luminosities for our sample (Table~\ref{tab:values}).  The exception is DM~Tau, for which we have no new X-ray observations to report.  

We note that most of our X-ray fluxes (Table~\ref{tab:xrayflux}) are consistent with previously published literature values within a factor of $\sim2$--3 \citep{gudel10,ingleby11a}. The exceptions are GM~Aur E5 and CS Cha.  Our GM~Aur E5 flux is $\sim8$ times higher than that found by \citet{gudel10}, but all other epochs of GM~Aur are consistent within a factor of two; as we discuss below, our findings support that GM~Aur was in a flaring state in E5.  Our CS~Cha flux is $\sim9$ times higher than previously reported \citep[$1.0\times10^{-12}$ erg cm$^{-2}$ s$^{-1}$;][]{gudel10}.  This suggests that we caught CS~Cha in a higher X-ray state.  We leave it to future work to explore this further with more epochs of data.

\subsection{The FUV \h2 Bump} \label{sec:h2}

The FUV continuum emission of CTTS can be explained by a combination of accretion shock emission 
\citep[e.g.,][]{ingleby15} and molecular gas emission that is dominated by \h2 \citep{herczeg02,herczeg04,bergin04,ingleby09,france11a,france11b}. Here we focus on the broad molecular and continuum feature at 1600~{\AA} known as the \h2 bump \citep[e.g.,][]{herczeg04,bergin04} and measure the luminosity of this feature.   

\begin{figure*}
\center
\subfloat{
\includegraphics[width=0.4\linewidth]{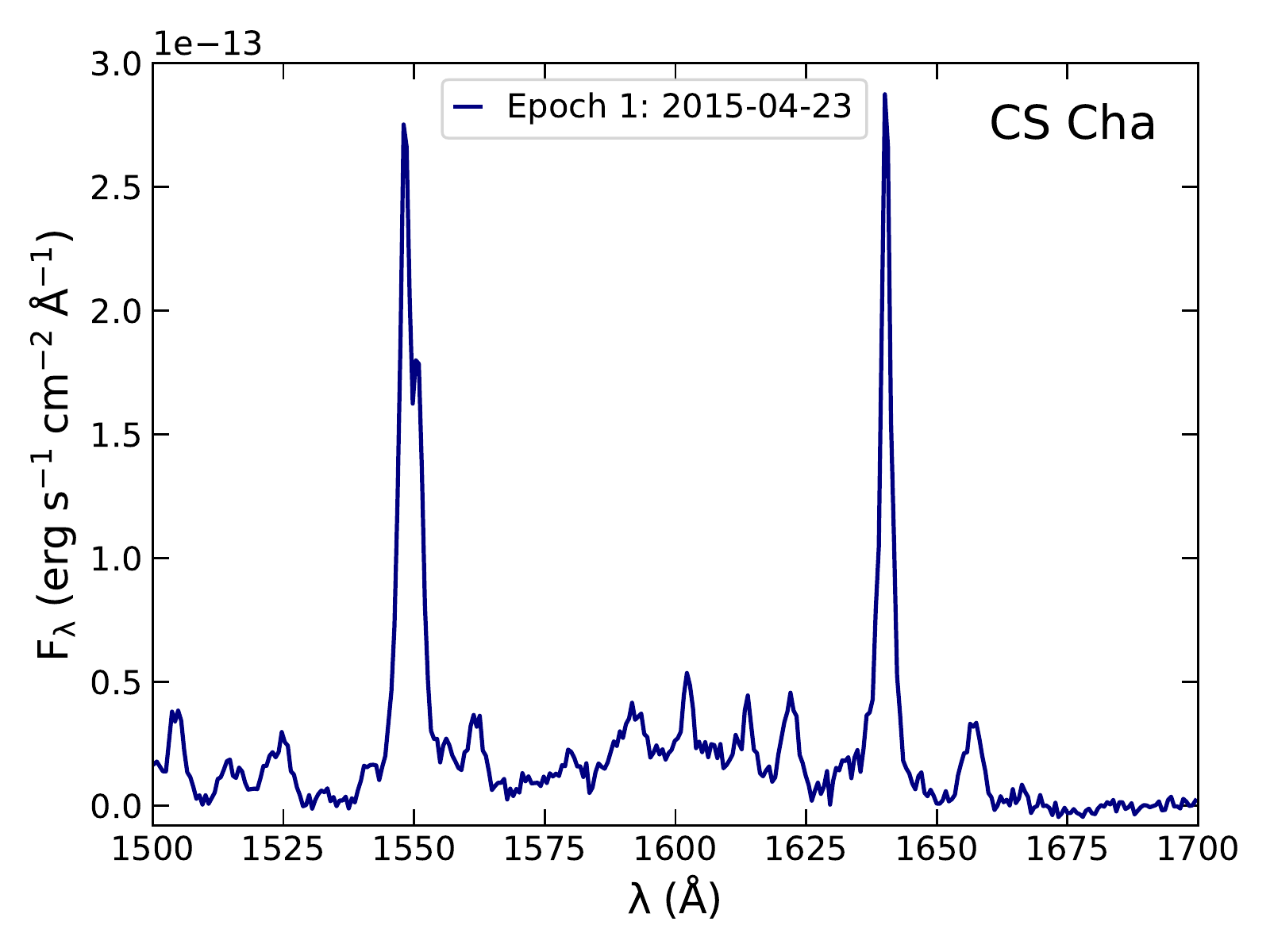}
}
\subfloat{
\includegraphics[width=0.4\linewidth]{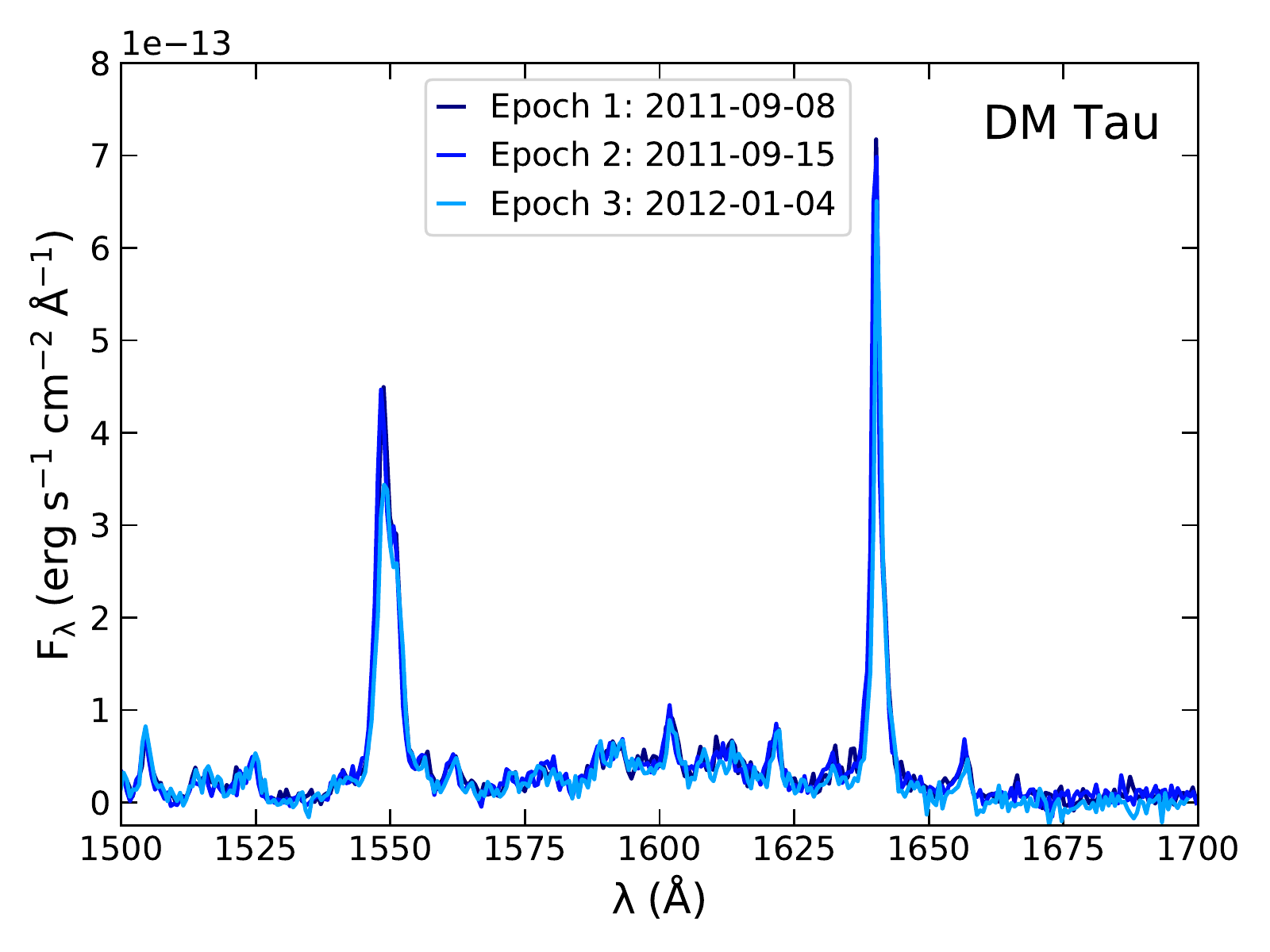}
}

\subfloat{
\includegraphics[width=0.4\linewidth]{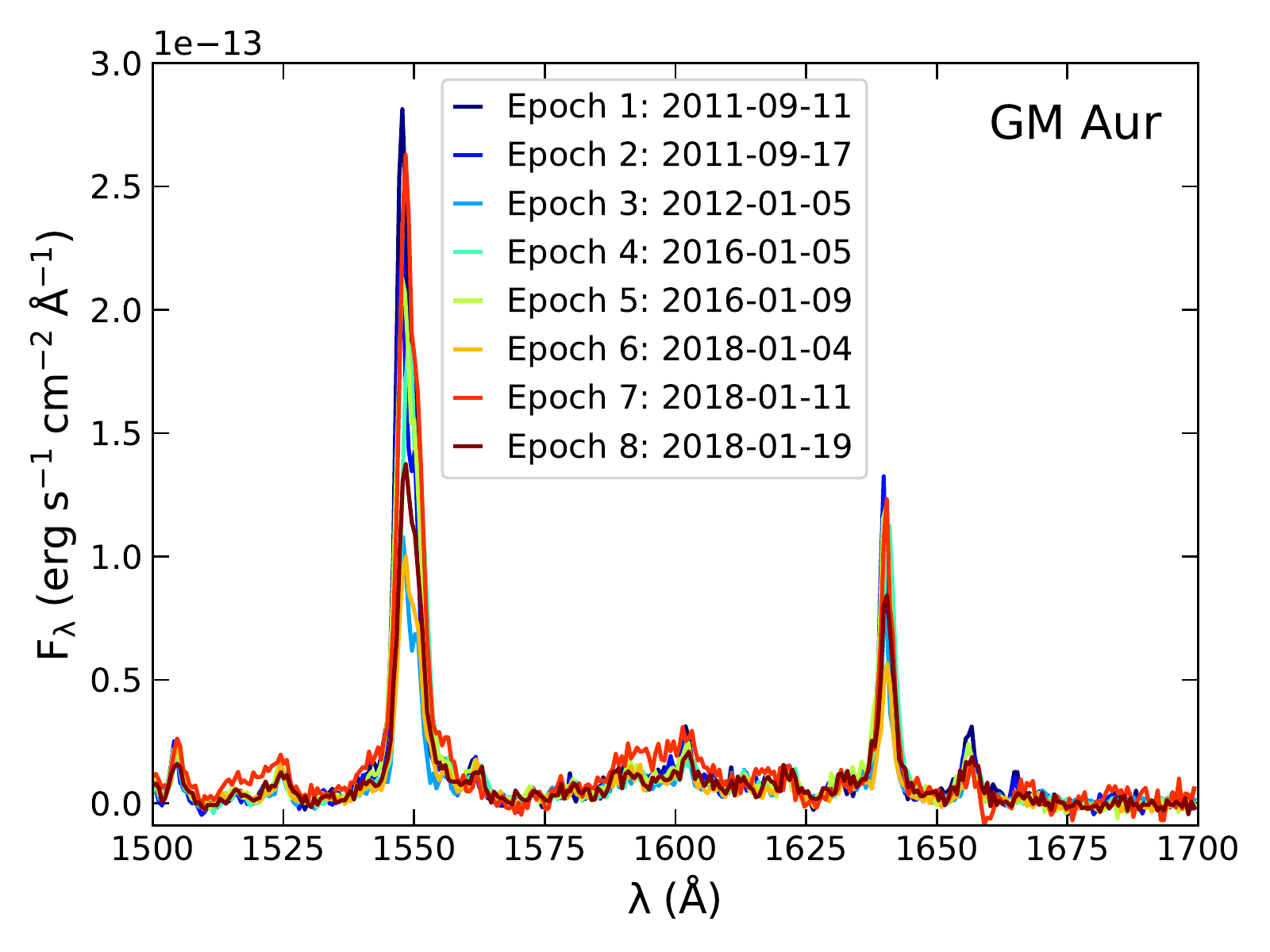}
}
\subfloat{
\includegraphics[width=0.4\linewidth]{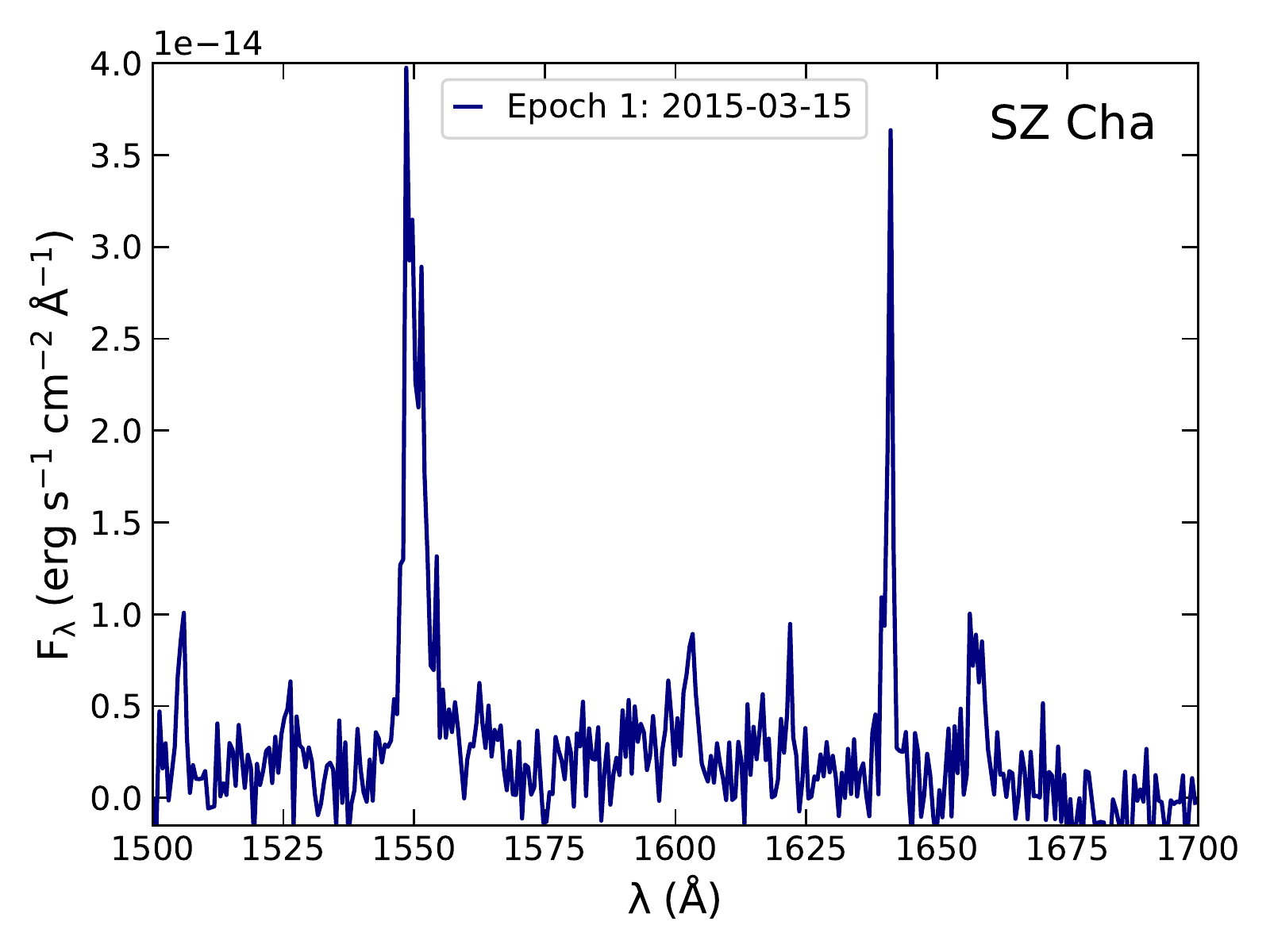}
}

\subfloat{
\includegraphics[width=0.4\linewidth]{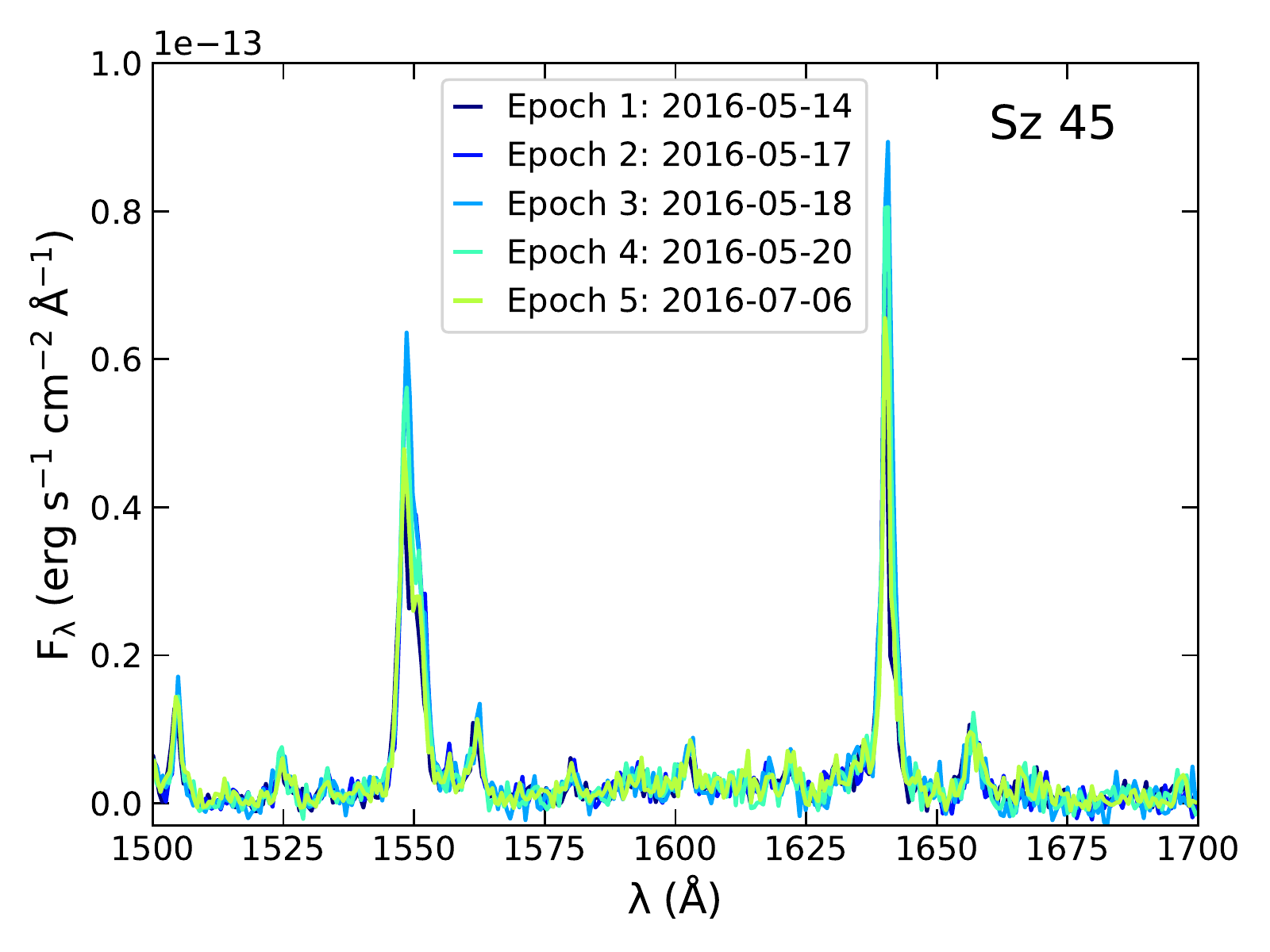}
}
\subfloat{
\includegraphics[width=0.4\linewidth]{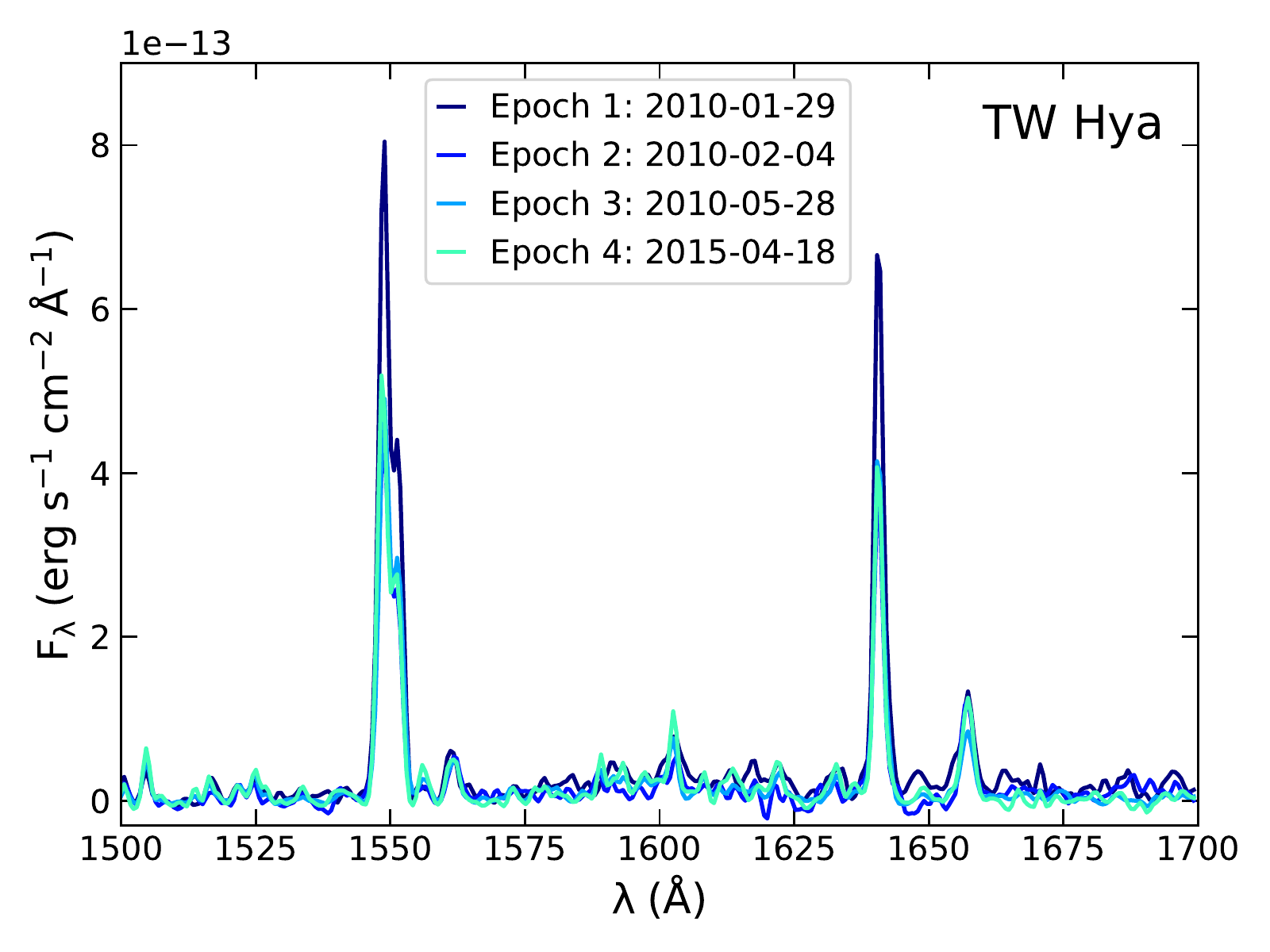}
}

\subfloat{
\includegraphics[width=0.4\linewidth]{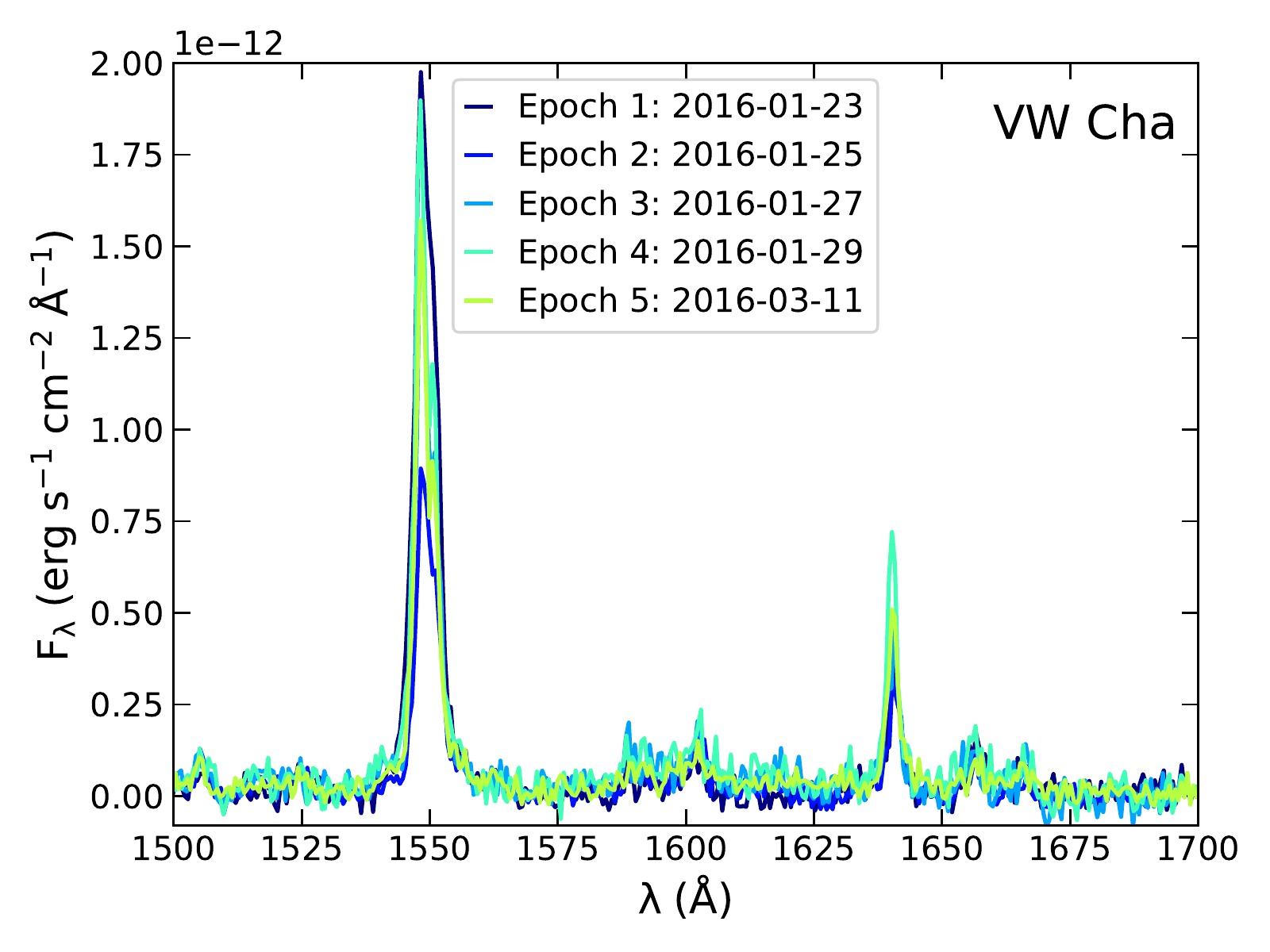}
}
\caption[]{ 
Continuum-subtracted {\it HST} FUV spectra of our sample centered on the \h2 bump at 1600~{\AA}. The measured \h2 bump luminosities are listed in Table~\ref{tab:values}.  Note that the strong lines at 1548~{\AA} and 1640~{\AA} are $C_{IV}$ and $He_{II}$, respectively.
}
\label{fig:h2}
\end{figure*}

In Figure~\ref{fig:h2}, we show the continuum-subtracted FUV emission of all objects in our sample.  GM Aur was likely undergoing an accretion burst in E7 (RE19), and we will return to this point in Section 4.  We dereddened all the {\it HST} FUV data using the A$_{V}$ values listed in Table~\ref{tab:starparam} and the extinction law toward HD~29647 \citep{whittet04}.  The underlying FUV continuum emission was removed using a third-degree polynomial fit to hand-selected points representative of the continuum level. The posterior for the fit was found using standard linear regression techniques. 

As noted earlier, the feature at 1600~{\AA} is a combination of the \h2 bump and Ly$\alpha$-fluoresced \h2. There are different methods for measuring the \h2 bump luminosity, depending on the resolution of the data.  For example, \citet{ingleby09} had low-resolution Advanced Camera for Surveys (ACS) spectra, and so their \h2 bump luminosity measurement had a combination of the \h2 bump and Ly$\alpha$-fluoresced \h2.  Meanwhile, \citet{france17} had much higher resolution Cosmic Origins Spectrograph (COS) spectra and excluded \h2 lines, leaving behind a more clean measurement of the \h2 bump.  While here we have lower resolution than COS, we are able to remove the strongest \h2 lines.  Therefore, our STIS-derived \h2 bump luminosities can be compared to those measured with COS with the caveat that there is likely still some Ly$\alpha$-fluoresced \h2 line emission.

Several of the strongest fluorescent \h2 emission lines were removed by hand using Gaussian line profiles. The lines that were removed (when present) include the following \h2 transitions: 3-9 R(15), 3-9 P(17), 3-10 R(15), 4-11 R(3), 4-11 P(5), 1-8 P(8), 1-9 R(3), 1-9 P(5), 3-10 P(1), 2-8 P(13), 2-9 R(11) \citep[see][]{herczeg06}. We then integrated the \h2 bump between 1570~{\AA} and 1630~{\AA}, avoiding the strong emission lines of $C_{IV}$, $C_{I}$, and $He_{II}$ at 1548~{\AA}, 1560~{\AA}, and 1640~{\AA}, respectively. A posterior for the \h2 bump luminosities was derived using a Markov chain Monte Carlo (MCMC) approach assuming Gaussian measurement uncertainties for the data (following RE19).  Measured values (Table~\ref{tab:values}) are derived from the 50th percentile of the \h2 bump luminosity posterior distribution, while reported uncertainties are 16th and 84th percentile values. We refer the reader to RE19 for analysis of other FUV lines. 

Previously reported measurements of the \h2 bump luminosity from high-resolution COS spectra are given by \citet{france17} for
CS~Cha ($6.99\pm1.70\times10^{29}$ erg s$^{-1}$), 
DM~Tau ($8.37\pm2.33\times10^{29}$ erg s$^{-1}$), 
GM~Aur ($20.27\pm5.47\times10^{29}$ erg s$^{-1}$), and 
TW~Hya ($8.49\pm1.71\times10^{29}$ erg s$^{-1}$).  
(We note that these have been scaled to the \textit{Gaia} distances listed in Table~\ref{tab:starparam}.)
Our measured \h2 bump luminosity values are roughly consistent within the measurement uncertainties
for GM~Aur and TW~Hya.  Our values for CS~Cha and DM~Tau are about 6--7 times higher than those of \citet{france17}.
Given the variable nature of these objects, we cannot determine whether this is due to intrinsic variability or whether the STIS resolution leads to overestimating the  \h2 bump luminosity in some cases but not others.  
In addition, line luminosity measurements depend on the adopted extinctions and, to some degree, the adopted spectral types.  Interestingly, we adopt the same extinction and spectral type as \citet{france17} for CS~Cha.  For DM~Tau, we adopt a different A$_{V}$ (= 1.1) than \citet[A$_{V}=0$;][]{france17}. We leave further exploration of whether this is due to intrinsic variability to future work.

To the best of our knowledge, SZ~Cha, Sz~45, and VW~Cha have no previously reported \h2 bump luminosities. The average \h2 bump luminosity in CTTS is $1\times10^{29}$ erg s$^{-1}$ \citep{france17}. SZ~Cha and Sz~45 are about 3--6 times higher, while VW~Cha is 50--100 times higher. As mentioned above, these higher-than-average measurements may be due to intrinsic variability or to adoption of an inappropriate A$_{V}$ and/or spectral type.

\subsection{Correlations} \label{sec:corr}

Here we search for correlations in our dataset between the luminosity of the \h2 bump and L$_X$ or L$_{acc}$.  We also search for correlations between L$_X$ and L$_{acc}$, \mdot, UV emission lines, or accretion column properties. To facilitate comparison with previous works, we report the Pearson correlation coefficient ($\rho_p$) and its p-value ($p_p$), the Spearman correlation coefficient ($\rho_s$) and its p-value ($p_s$), and the Kendall correlation coefficient ($\tau_k$) and its p-value ($p_k$).

\begin{figure*}
\epsscale{1.0}
\plottwo{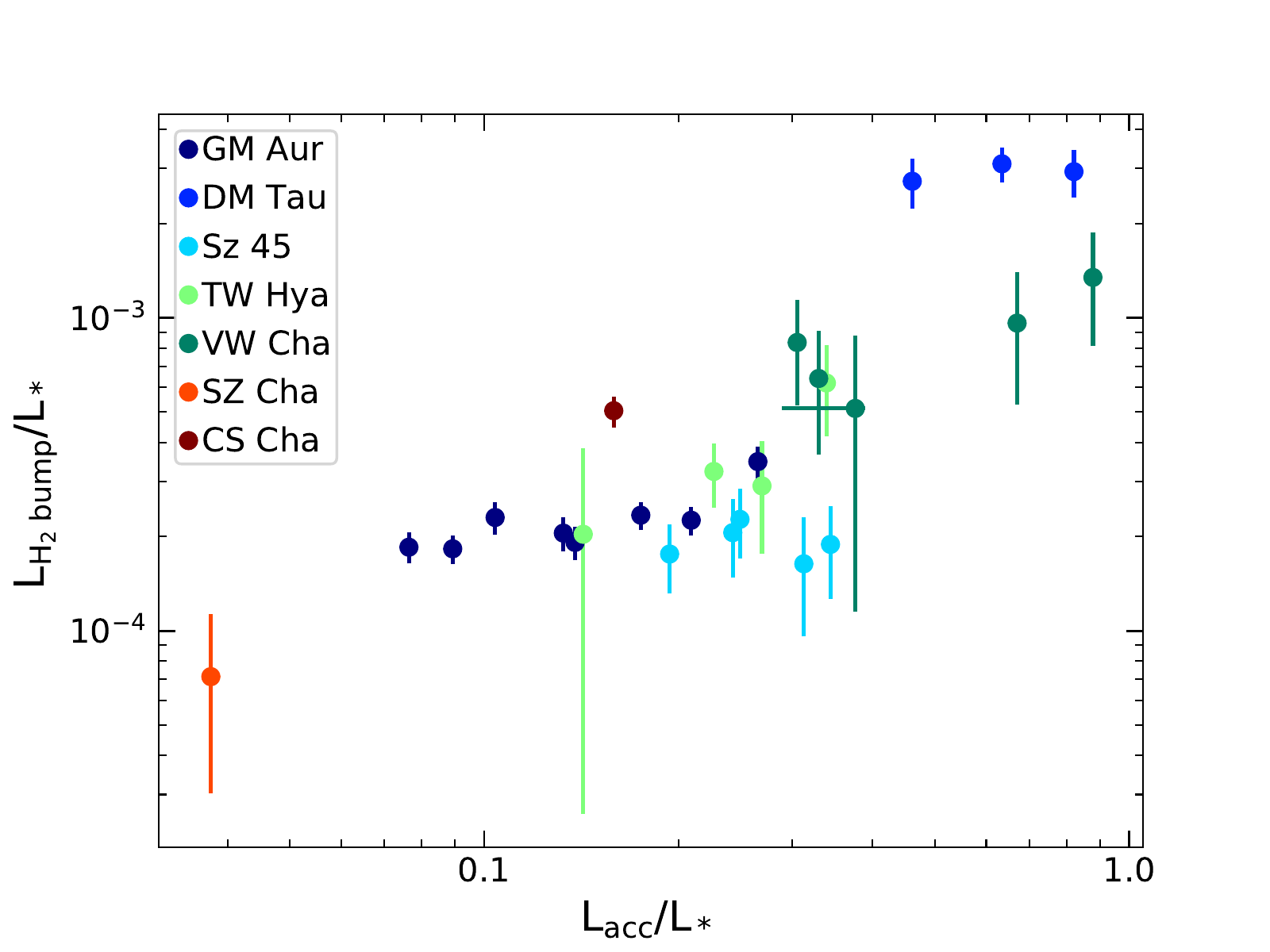}{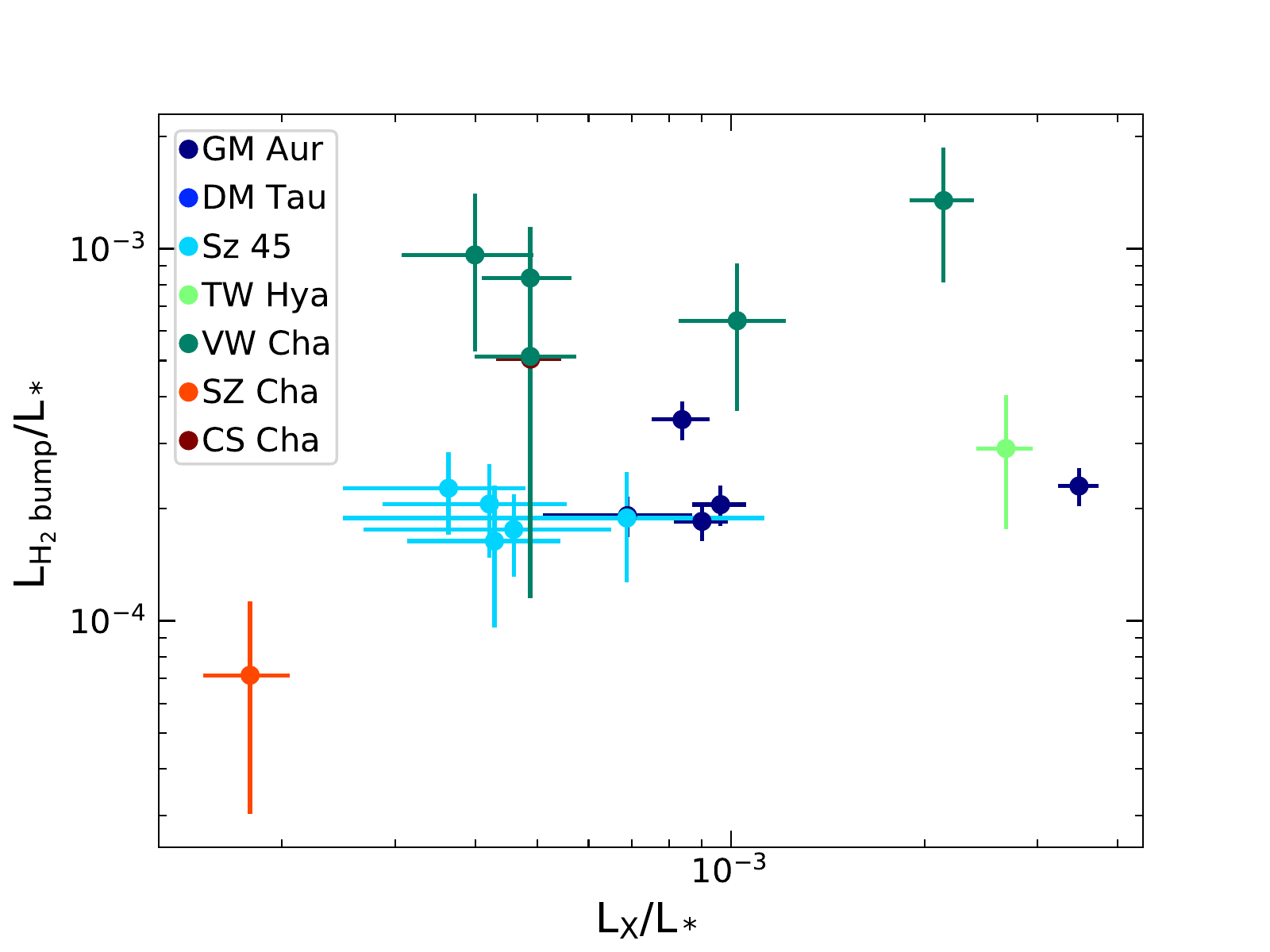}
\caption[]{
Left: Comparison of the H$_2$ bump luminosity to the accretion luminosity, L$_{acc}$. There is a correlation ($\rho_p$=0.8, $p_p$=4e-6; $\rho_s$=0.7, $p_s$=1e-5; $\tau_k$=0.6, $p_k$=3e-5). Right: Comparison of the H$_2$ bump luminosity to the X-ray luminosity, L$_X$. There is no correlation ($\rho_p$=0.1, $p_p$=0.6; $\rho_s$=0.3, $p_s$=0.3; $\tau_k$=0.2, $p_k$=0.3).
}
\label{fig:h2lacc}
\label{fig:h2lx}
\end{figure*}

We find a positive correlation between the \h2 bump luminosity and L$_{acc}$ ($\rho_p$=0.8, $p_p$=4e-6; $\rho_s$=0.7, $p_s$=1e-5; $\tau_k$=0.6, $p_k$=3e-5). In Figure~\ref{fig:h2lacc} (left), we plot the \h2 bump luminosity compared to L$_{acc}$.  
DM~Tau is offset from the rest of the sample, and removing it has an unclear effect on the correlation 
($\rho_p$=0.9, $p_p$=5e-8; $\rho_s$=0.6, $p_s$=8e-4; $\tau_k$=0.5, $p_k$=6e-4).  In accretion-shock model fitting of DM~Tau, RE19 found that DM~Tau had more excess at the shortest NUV wavelengths relative to the rest of the sample, and an additional higher-energy ($\curf=3\times10^{12}$ erg s$^{-1}$ cm$^{-3}$) accretion column best reproduced the data.  However, it is unclear how or if this affects the correlation seen here, and we leave it for future work to explore this further.  
If we remove both DM~Tau and the two epochs of VW~Cha with the highest L$_{acc}$ (E1 and E2), the correlation between the \h2 bump luminosity and L$_{acc}$ is weaker ($\rho_p$=0.6, $p_p$=5e-3; $\rho_s$=0.5, $p_s$=0.01; $\tau_k$=0.4, $p_k$=9e-3). 
Other works \citep{ingleby09,france17} have found positive correlations between the \h2 bump luminosity and L$_{acc}$. We discuss the implications of this correlation between the \h2 bump luminosity and L$_{acc}$ further in Section~\ref{sec:dis-h2}.
 
\begin{figure*}
\center
\subfloat{
\includegraphics[width=0.4\linewidth]{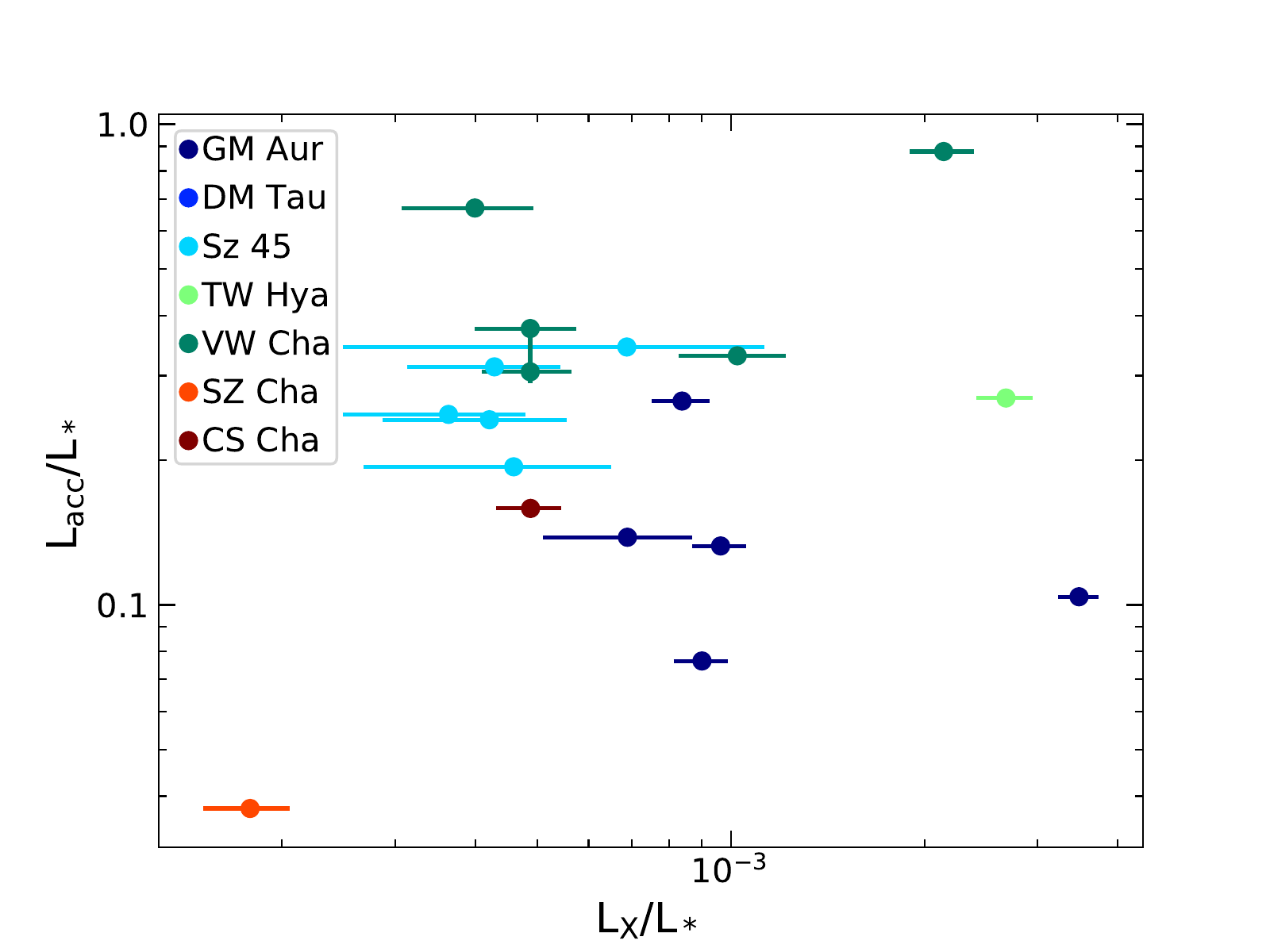}
}
\subfloat{
\includegraphics[width=0.4\linewidth]{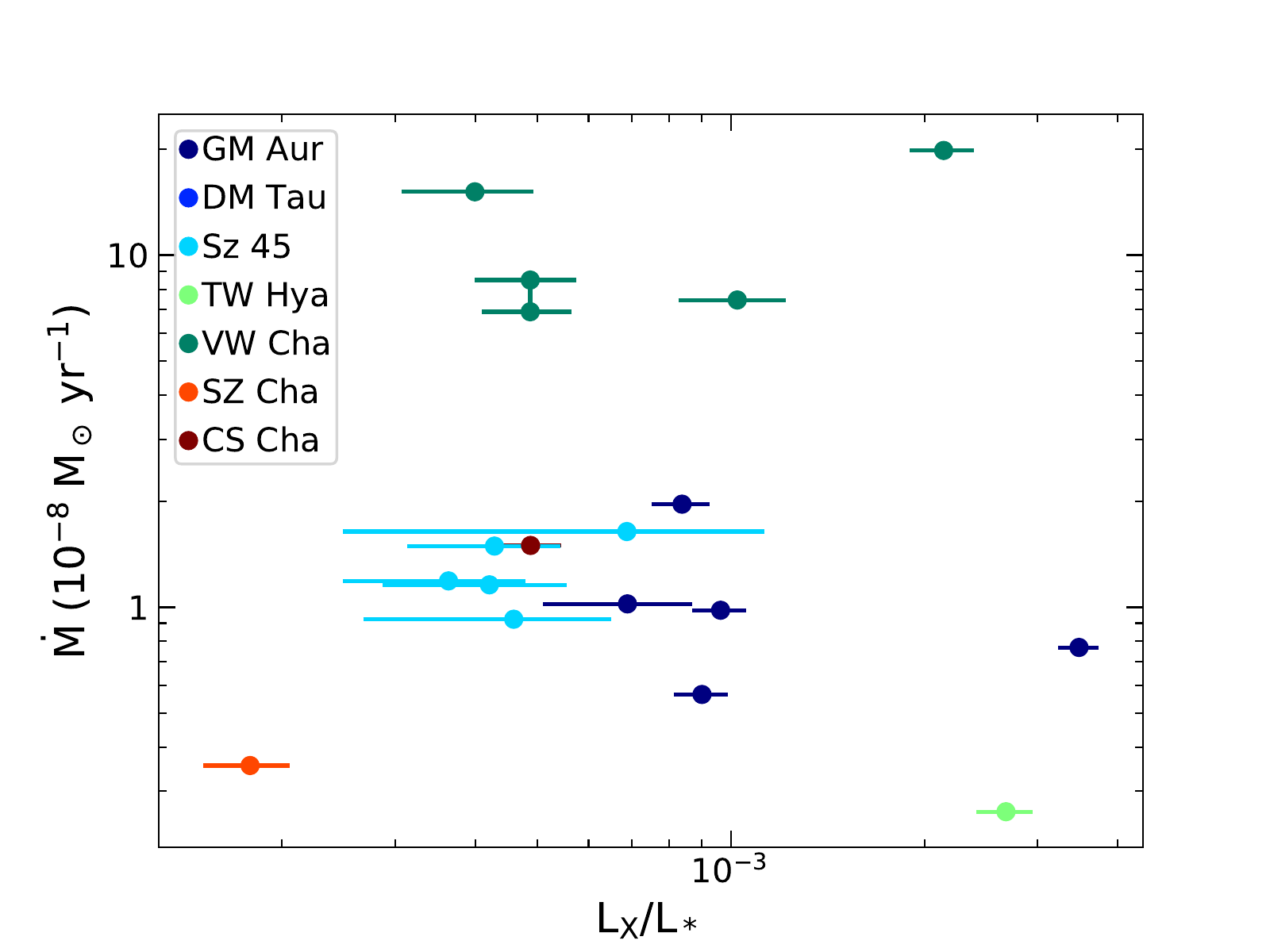}
}

\subfloat{
\includegraphics[width=0.4\linewidth]{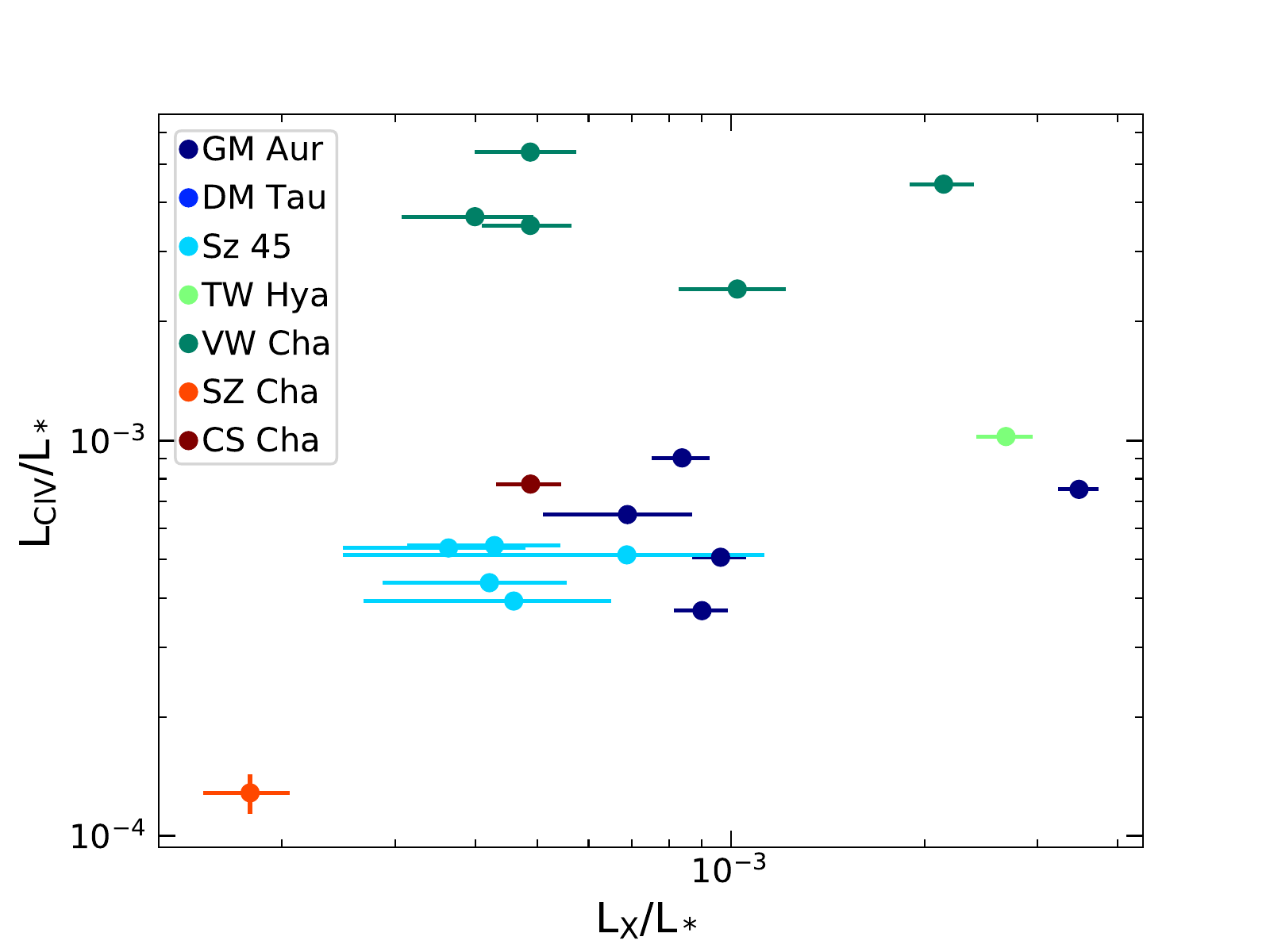}
}
\subfloat{
\includegraphics[width=0.4\linewidth]{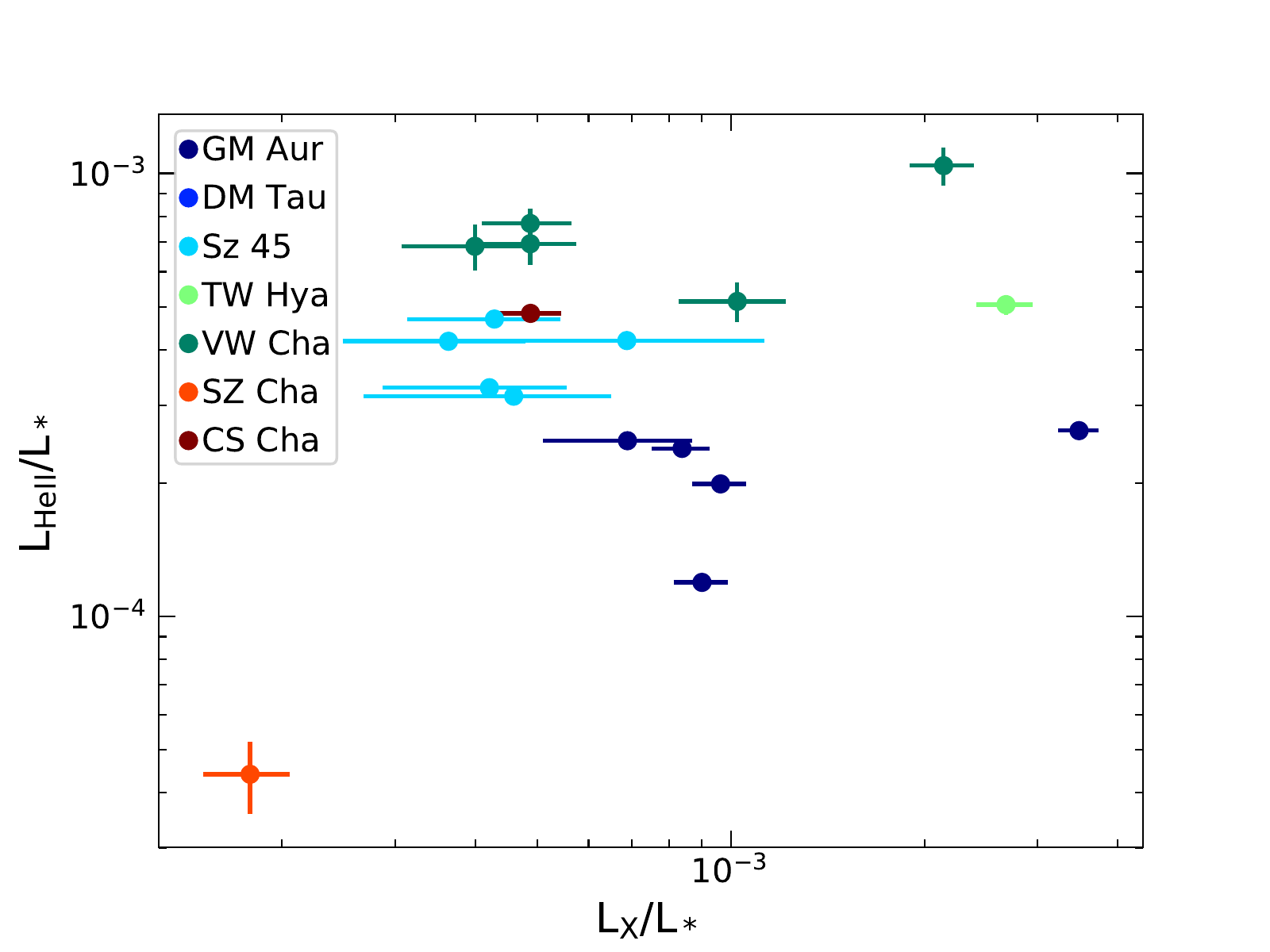}
}

\subfloat{
\includegraphics[width=0.4\linewidth]{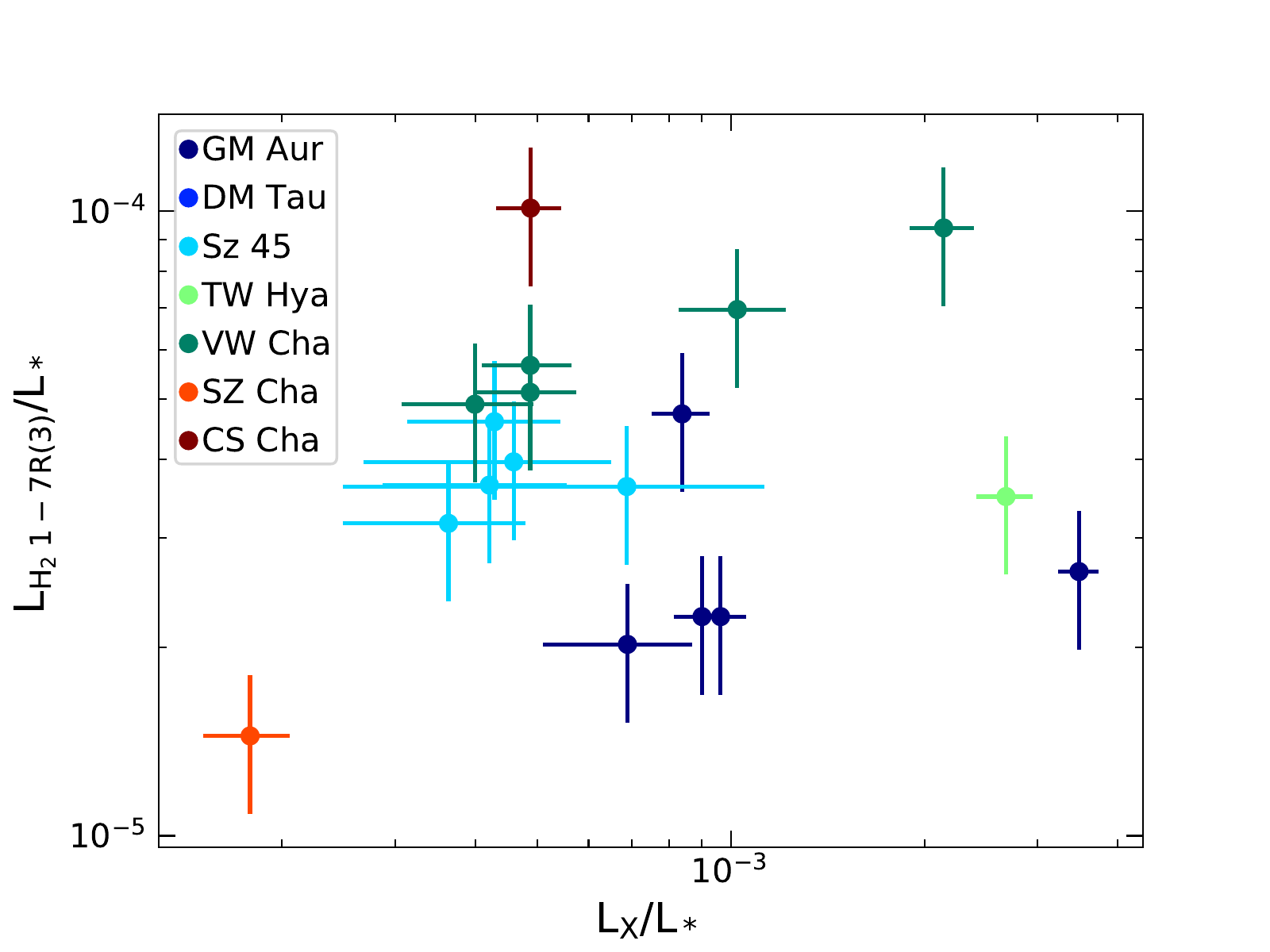}
}
\subfloat{
\includegraphics[width=0.4\linewidth]{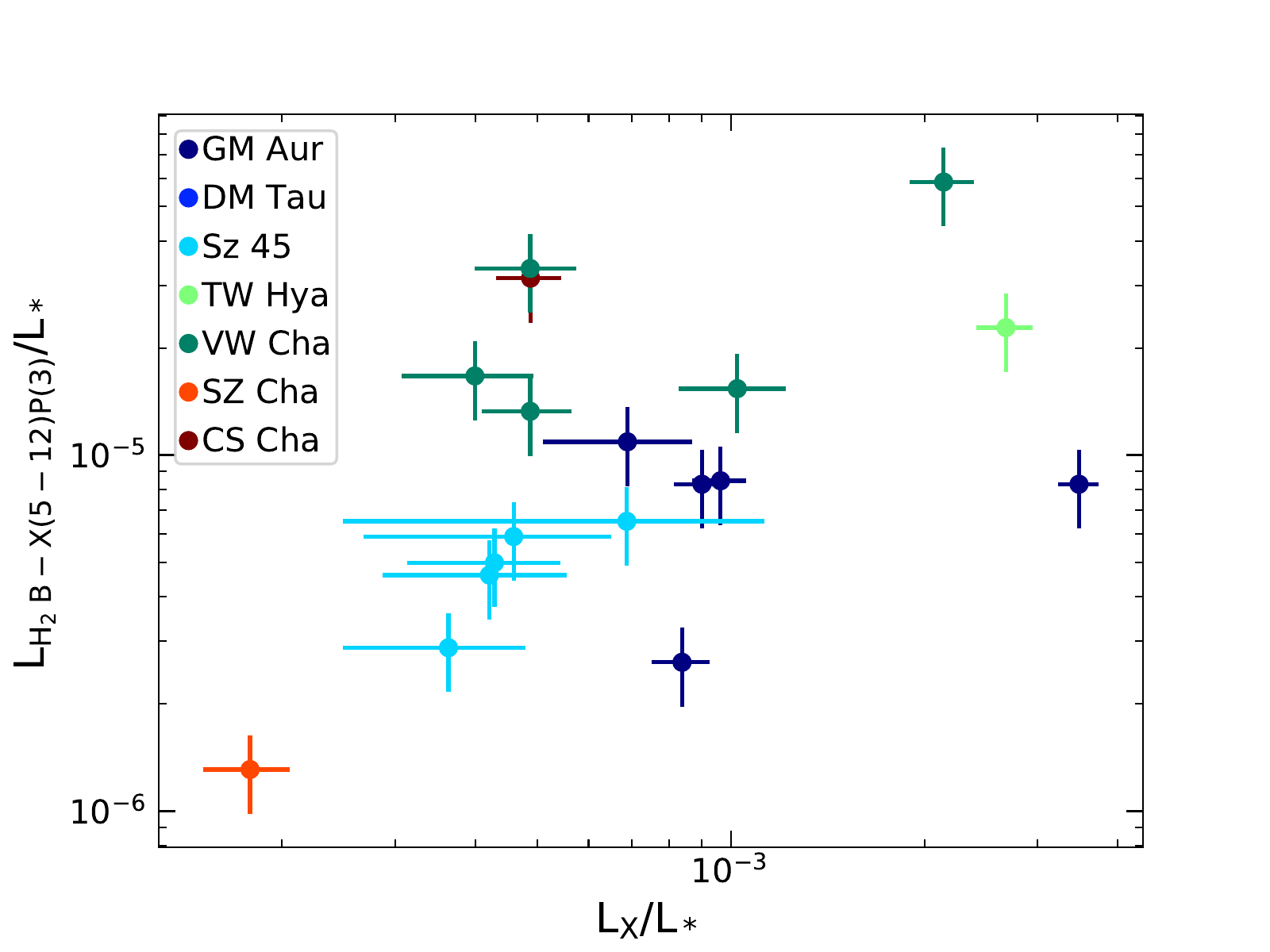}
}
\caption[]{Comparison of the X-ray luminosity, L$_X$, to the accretion luminosity, L$_{acc}$ (top left), the accretion rate, \mdot\ (top right), and the line luminosities of
$C_{IV}$ at $1548~{\AA}$ (middle left),
$He_{II}$ at $1640~{\AA}$ (middle right),
\h2 1-7R(3) at $1489.5~{\AA}$ (bottom left), 
and \h2 B-X(5-12)P(3) at $1613~{\AA}$ (bottom right).  
There are no detected correlations with L$_X$ (see Section~\ref{sec:corr}).
}
\label{fig:lacclx}
\end{figure*} 

We find no significant correlations in our sample between L$_X$ and the \h2 bump luminosity, L$_{acc}$, or \mdot. There is no correlation between L$_X$ and the \h2 bump luminosity ($\rho_p$=0.1, $p_p$=0.6; $\rho_s$=0.3, $p_s$=0.3; $\tau_k$=0.2, $p_k$=0.3; Figure~\ref{fig:h2lacc}, right). \citet{france17} also reported the lack of a correlation between L$_X$ and the \h2 bump luminosity.  We also find no correlation between L$_X$ and L$_{acc}$ ($\rho_p$=0.1, $p_p$=0.7; $\rho_s$=--0.01, $p_s$=0.96; $\tau_k$=--0.1, $p_k$=0.8; Figure~\ref{fig:lacclx}, top left) or \mdot\ ($\rho_p$=0.1, $p_p$=0.8; $\rho_s$=--0.1, $p_s$=0.7, $\tau_k$=--0.04, $p_k$=0.82; Figure~\ref{fig:lacclx}, top right). We discuss the implications of the lack of correlation between L$_X$ and the \h2 bump luminosity in Section~\ref{sec:dis-h2} and between L$_X$ and L$_{acc}$ or \mdot\ in Section~\ref{sec:dis-xray}.  

We also searched for correlations between L$_X$ and emission lines measured in the STIS data by RE19.  The lines we investigated from Table~7 of RE19 are as follows: $C_{II}$ $1335~{\AA}$,  $C_{I}$ $1463~{\AA}$,
$C_{IV}$ $1548~{\AA}$,  $He_{II}$ $1640~{\AA}$, $O_{III}$ $1666~{\AA}$, 
$Si_{II}$ $1808~{\AA}$, $Si_{III]}$ $1892~{\AA}$,  $C_{III]}$ $1908~{\AA}$,  $C_{II]}$ $2325~{\AA}$,  $Al_{III]}$ $2670~{\AA}$,  
$Mg_{II}$ $2796~{\AA}$. We find no correlations with L$_X$.  In Figure~\ref{fig:lacclx}, we show comparisons between L$_X$ and $C_{IV}$ (middle left) and $He_{II}$ (middle right), the two strongest lines in Figure~\ref{fig:h2}.    We find no correlation between $C_{IV}$ and L$_X$ ($\rho_p$=0.04, $p_p$=0.9; $\rho_s$=0.3, $p_s$=0.2; $\tau_k$=0.2, $p_k$=0.3) or $He_{II}$ and L$_X$ ($\rho_p$=0.1, $p_p$=0.6; $\rho_s$=0.05, $p_s$=0.9; $\tau_k$=0.0, $p_k$=1.0). We note that $He_{II}$ has been previously linked to X-ray emission \citep{alexander05}. We also measure two Lyman-band \h2 transition lines (Table~\ref{tab:values}) from \citet{herczeg06} and compare them to L$_X$ in Figure~\ref{fig:lacclx}. We find no correlation with  \h2 1-7R(3) at $1489.5~{\AA}$ (bottom left, $\rho_p$=0.02, $p_p$=0.9; $\rho_s$=0.04, $p_s$=0.9; $\tau_k$=0.08, $p_k$=0.6) or \h2 B-X(5-12)P(3) at $1613~{\AA}$ (bottom right, $\rho_p$=0.3, $p_p$=0.2; $\rho_s$=0.5, $p_s$=0.1; $\tau_k$=0.4, $p_k$=0.04).

\begin{figure}
\epsscale{1.0}
\plotone{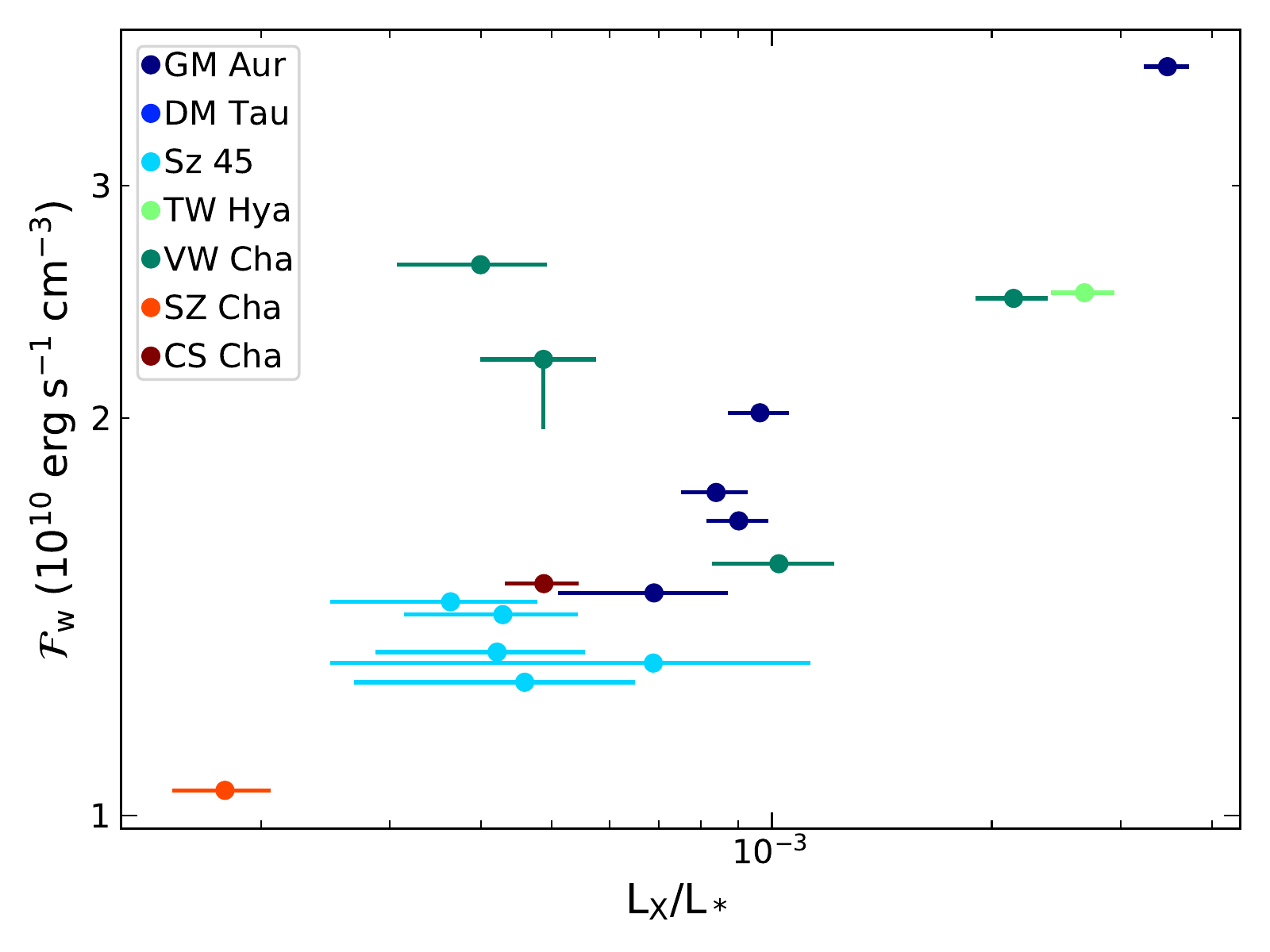}
\caption[]{
Comparison of the weighted energy flux, ${\curf}_w$, of the accretion columns to the X-ray luminosity, L$_X$. There is a correlation ($\rho_p$=0.8, $p_p$=9e-5; $\rho_s$=0.6, $p_s$=0.009; $\tau_k$=0.5, $p_k$=0.005). }
\label{fig:curf}
\end{figure} 

Interestingly, we do see a correlation between the accretion column energy flux and L$_{X}$ ($\rho_p$=0.8, $p_p$=9e-5; $\rho_s$=0.6, $p_s$=0.009; $\tau_k$=0.5, $p_k$=0.005). The shock models we used to measure \mdot\ consist of three columns with an energy flux, $\curf$ (= $1\times10^{10}$, $1\times10^{11}$, $1\times10^{12}$ erg s$^{-1}$ cm$^{-3}$), and a surface filling factor for each column, $f$ (Table~4 of RE19 and Table~6 in the Appendix). In Figure~\ref{fig:curf}, we plot the average energy flux weighted by the filling factor of each column, ${\curf}_w$, against L$_X$. We did not include VW Cha E5 since in this epoch, the accretion column may have obscured the stellar photosphere leading to optical dimming (RE19). If so, the accretion column may have absorbed some of the X-ray emission as well.  If we include VW Cha E5, the correlation between ${\curf}_w$ and L$_X$ is much weaker ($\rho_p$=0.1, $p_p$=0.7; $\rho_s$=0.5, $p_s$=0.03; $\tau_k$=0.4, $p_k$=0.02). This correlation between ${\curf}_w$ and L$_X$ is largely driven by GM~Aur E5, which had the highest $\curf_w$ ($\sim3\times10^{10}$ erg s$^{-1}$ cm$^{-3}$) and the highest L$_X$.  If we remove GM~Aur E5, the correlation is weaker
($\rho_p$=0.6, $p_p$=0.01; $\rho_s$=0.5, $p_s$=0.03; $\tau_k$=0.4, $p_k$=0.02).
We discuss this correlation between ${\curf}_w$ and L$_X$ further in Section~\ref{sec:dis-xray}.

\begin{figure}
\subfloat{
\includegraphics[width=\linewidth]{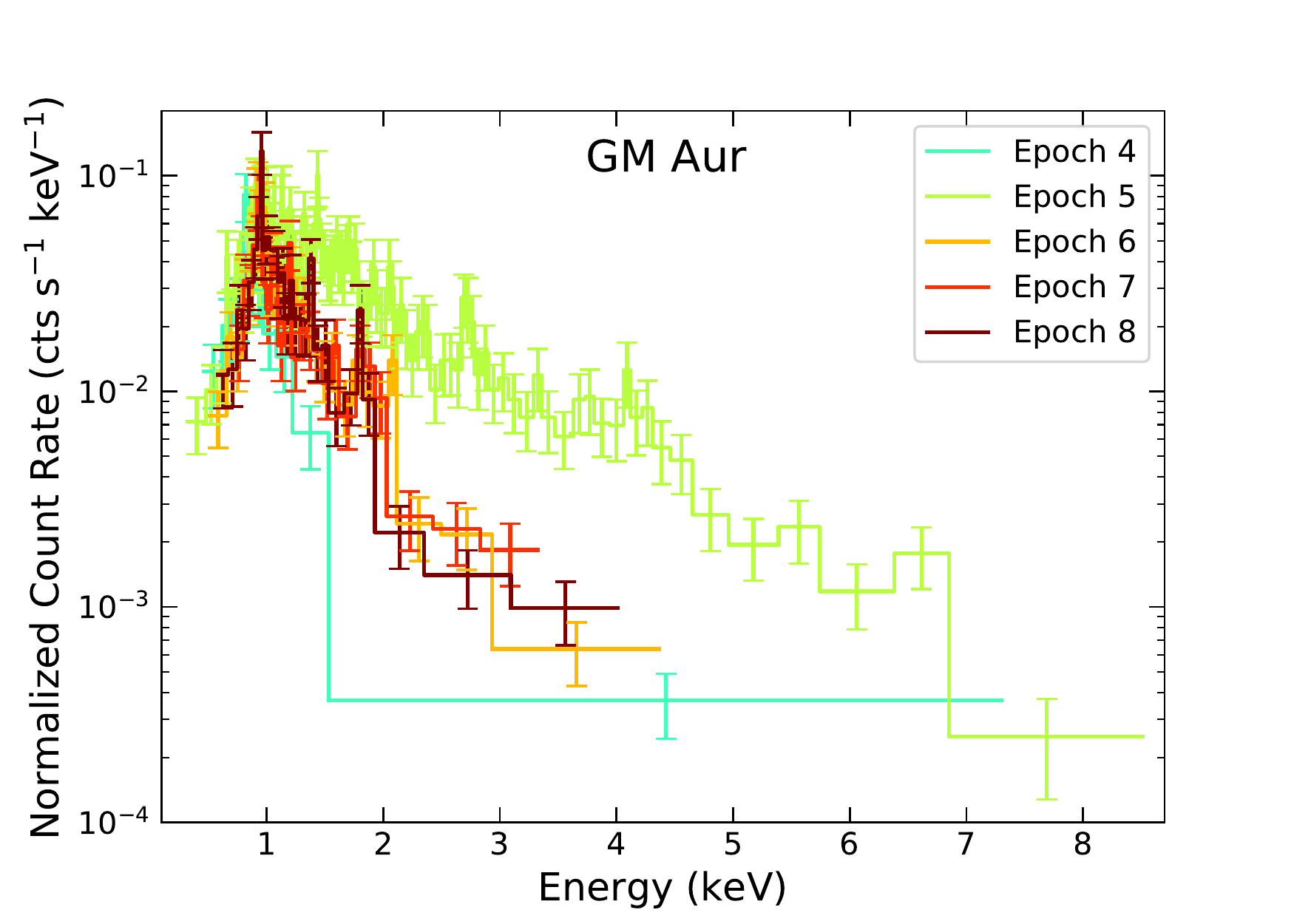}
}

\subfloat{
\includegraphics[width=\linewidth]{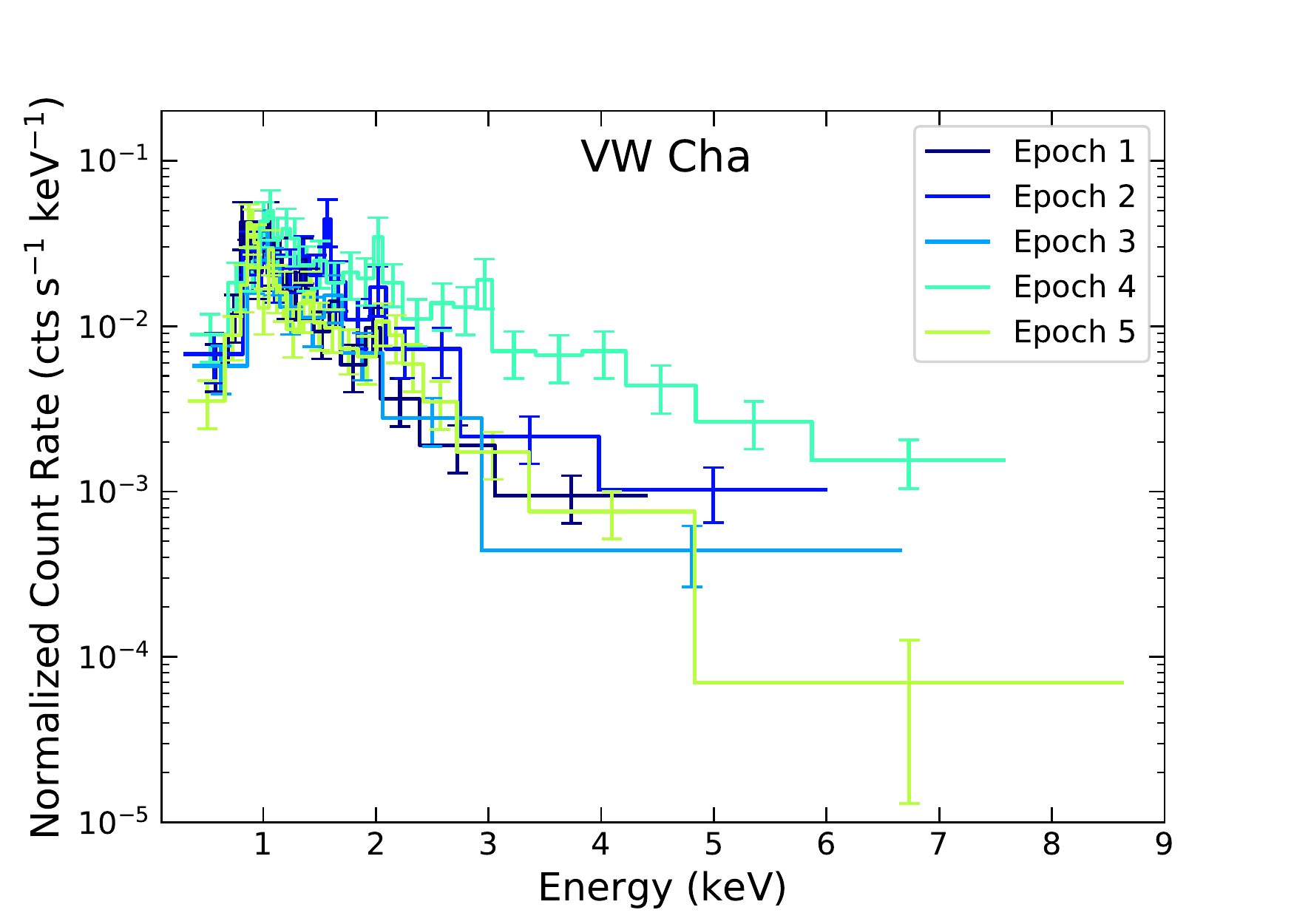}
}

\subfloat{
\includegraphics[width=\linewidth]{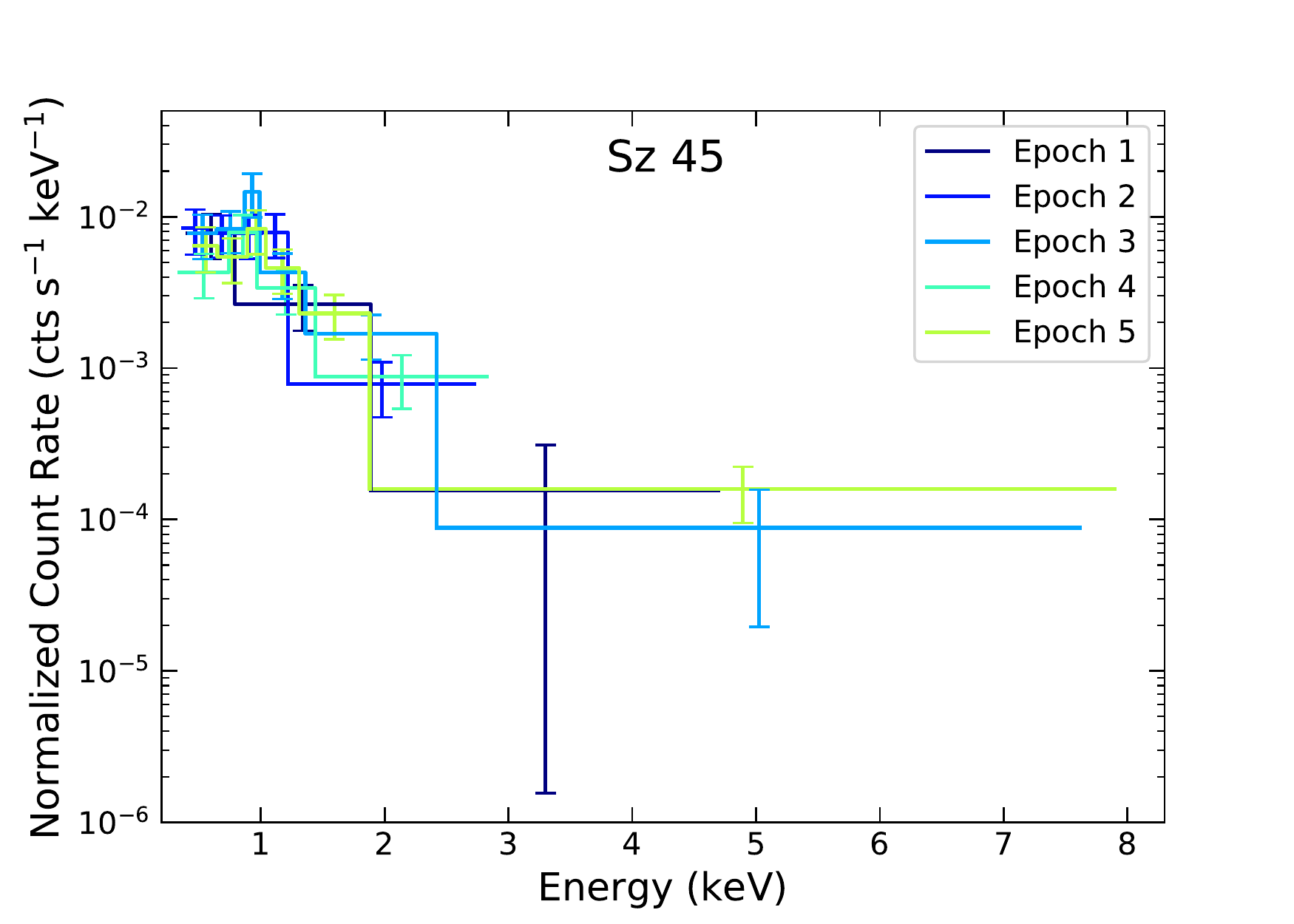}
}

\caption[]{
Multiepoch X-ray spectra for GM~Aur, VW~Cha, and Sz~45. GM~Aur E4 and E5 are {\it Swift} data.  GM~Aur E6, E7, and E8 are {\it Chandra} data. For VW Cha and Sz~45, we show {\it Swift} data. An increase in the hard X-ray emission is seen in GM~Aur E5 and VW~Cha E4, indicative of stellar flaring activity. No evidence of flaring activity is seen in Sz~45.  
}
\label{fig:flares}
\end{figure} 

Lastly, we investigated the X-ray spectra of objects with multiple epochs of X-ray data (GM~Aur, VW~Cha, and Sz~45; Figure~\ref{fig:flares}). In GM~Aur (top panel), we see that in E4, E6, E7, and E8, the X-ray spectra are very similar. However, in E5, the hard X-ray emission (1.5--8.0 keV) increases significantly while the soft X-ray emission stays the same. We note that between E4 and E5, \mdot\ did not change significantly (while L$_X$ did change), and in E7, there was a large increase in \mdot\ (while L$_X$ did not change). Although not as strong as seen in GM~Aur, we see a similar increase in the X-ray emission in E4 of VW~Cha (Figure~\ref{fig:flares}, middle panel) while the \mdot\ in this epoch was not significantly higher. In Sz~45, we see no evidence for significant changes in X-ray emission (Figure~\ref{fig:flares}, bottom panel).  The increase in the hard X-ray emission of GM~Aur and VW~Cha is indicative of stellar coronal X-ray flaring activity.  We discuss the above in light of the correlation between ${\curf}_w$ and L$_X$ in Section~\ref{sec:dis-xray}. 

\section{Discussion} \label{sec:dis} 
 
In this work, we find a correlation between the FUV \h2 bump luminosity and L$_{acc}$, but not L$_X$. We also see a correlation between L$_X$ and the density of the accretion column. Here we discuss the connection between the variability in the \h2 bump luminosity and L$_{acc}$ and the implications of our results on the connection between X-ray emission and accretion in TTS.

\begin{figure}
\subfloat{
\includegraphics[width=\linewidth]{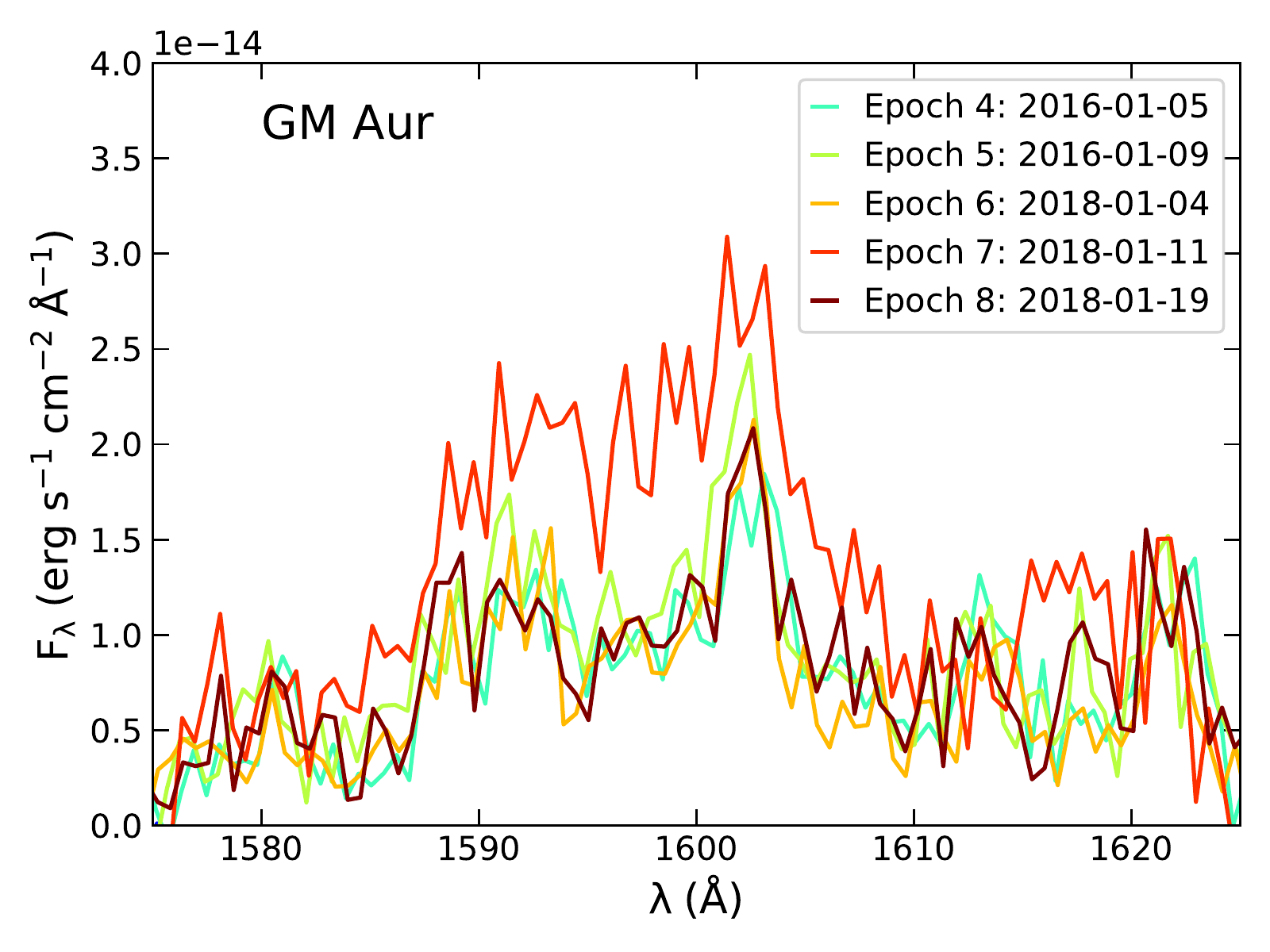}
}

\subfloat{
\includegraphics[width=\linewidth]{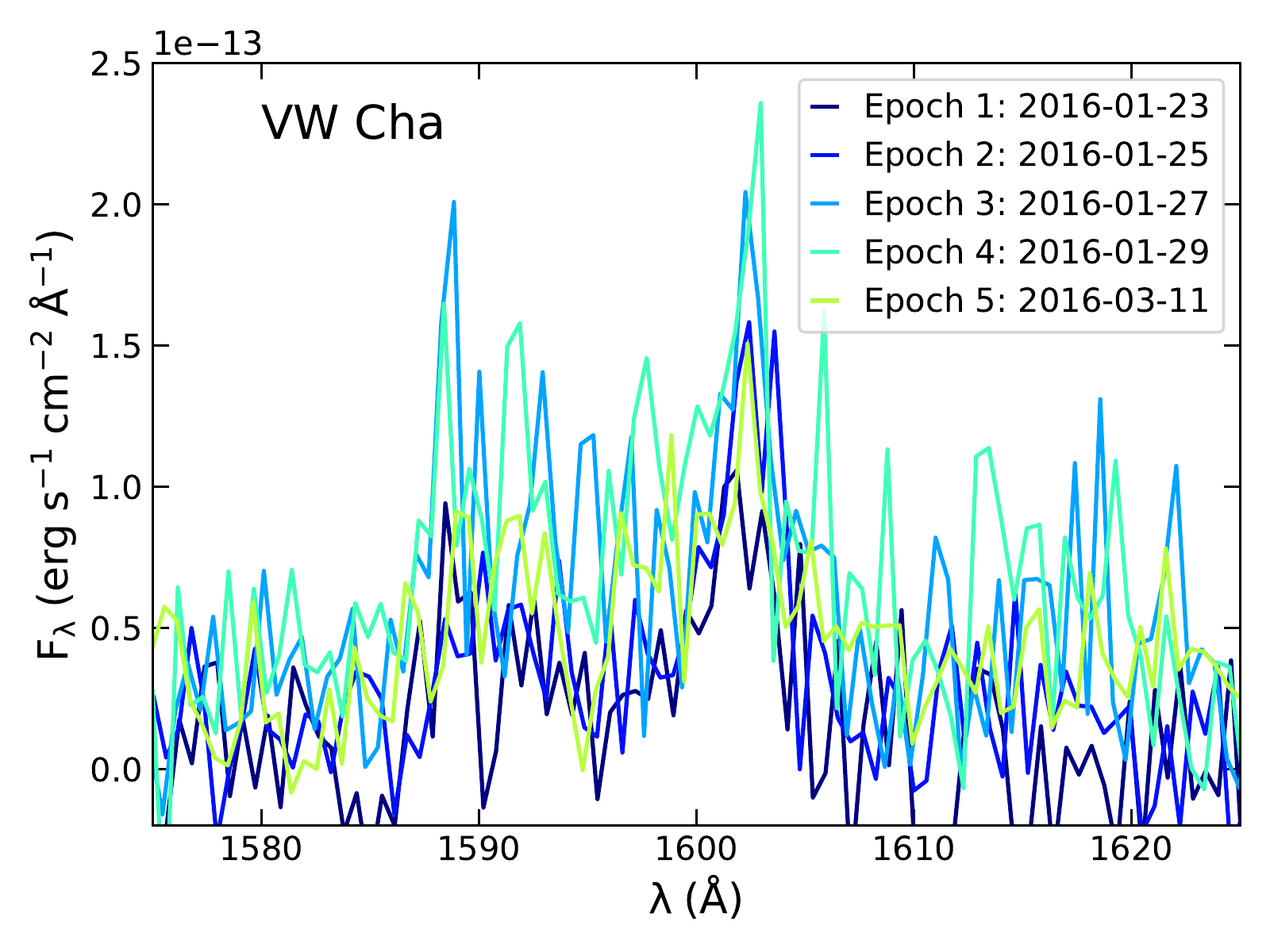}
}

\caption[]{ 
Continuum-subtracted {\it HST} FUV spectra of GM~Aur and VW~Cha from Figure~\ref{fig:h2} zoomed in on the \h2 bump at 1600~{\AA}. We only show those epochs of GM~Aur that have coordinated X-ray data.  We also note that stellar flaring activity is present in GM~Aur E5 and VW~Cha E4 (Figure~\ref{fig:flares}) and that there was an accretion burst in GM~Aur E7 (Table~\ref{tab:values}; RE19).
}
\label{fig:h2zoom}
\end{figure} 

\subsection{On the Origin of the FUV \h2 Bump} \label{sec:dis-h2}

\subsubsection{Previously Proposed Mechanisms:\\ X-ray vs.\ Ly$\alpha$ Emission}

\citet{herczeg04} and \citet{bergin04} proposed that the \h2 bump feature was a consequence of collisional excitation of \h2 by fast electrons in the inner disk kicked out from heavy elements by X-ray photons. Therefore, one would expect that this would lead to a correlation between the \h2 bump and X-ray luminosities. Here we find that the \h2 bump luminosity does not increase as L$_X$ increases in our sample (Figure~\ref{fig:h2lx}).
In Figure~\ref{fig:h2zoom}, we present the \h2 bump feature in GM~Aur and VW~Cha, which had stellar flaring in E5 and E4, respectively (Figure~\ref{fig:flares}).  The \h2 bump feature is not substantially higher in GM~Aur E5 and VW~Cha E4 relative to other epochs. However, the \h2 bump is much higher in GM~Aur E7, which we return to below in Section~\ref{subalt}.

We note that a correlation between the \h2 bump and X-ray luminosities has not been observed in large samples \citep{france17}.  In addition, using high-resolution COS FUV data, \citet{france11a} and \citet{france11b} noted that the \h2 bump is not centered near the expected 1575~{\AA} dissociation peak associated with electron-impact \h2. Also, the expected \h2 emission spectrum from electron-impact excitation was not seen \citep{france17}.  We do not have the resolution in our STIS data to robustly determine the peak of the \h2 bump nor the \h2 emission spectrum from electron-impact excitation.  However, our results support the intrepretation that X-ray ionization cannot be traced with the \h2 bump. 

\citet{france17} found a correlation between the \h2 bump luminosity and noncoordinated reconstructed/extrapolated Ly$\alpha$ fluxes and suggested that the \h2 bump was instead powered by Ly$\alpha$ photons, particularly Ly$\alpha$-driven dissociation of water in the inner disk. In this scenario, Ly$\alpha$ would be due to the strong stellar and accretion-generated Ly$\alpha$ radiation field. 
The accretion-related origin of Ly$\alpha$ is supported by modeling of H$\alpha$ and H$\beta$ emission lines that maps these lines to the accretion funnel flows \citep{alencar12}.
Excitation by Ly$\alpha$ photons would populate the upper levels of \h2, and a fluorescent spectrum would be emitted as it de-excites. This would make the \h2 bump an FUV spectral signature of \h2O dissociation, which has important implications on the water chemistry in the Ly$\alpha$-irradiated disk layers \citep{france17}.

\begin{figure}
\epsscale{1.2}
\plotone{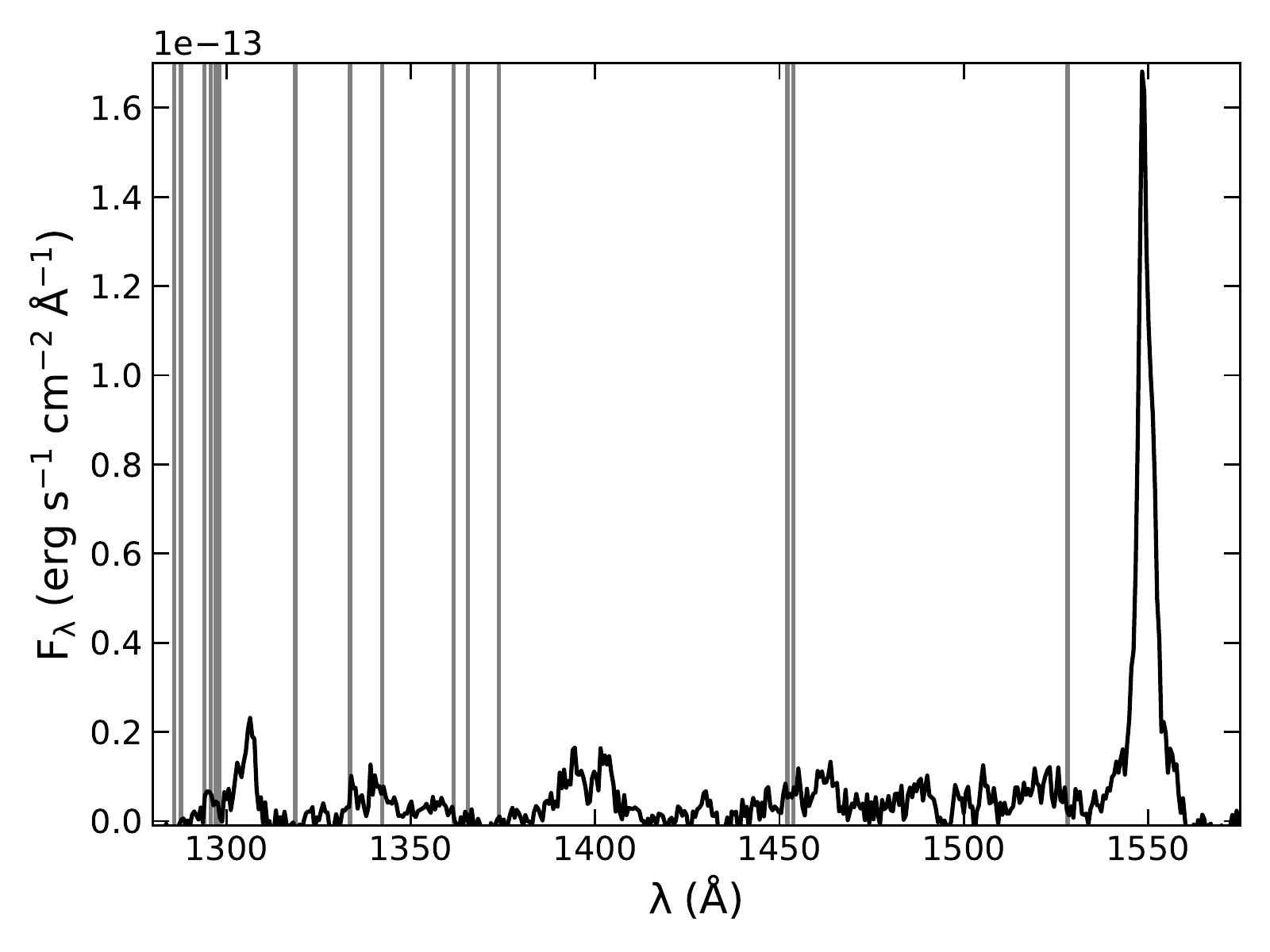}
\caption[]{
Excess emission in GM~Aur E7 relative to E3. We note that the spectra are continuum-subtracted, and we label the wavelengths corresponding to \h2 lines created by Ly$\alpha$ fluorescence in gray. There is no significant difference in the emission of these Ly$\alpha$-driven lines between the two epochs.
}
\label{fig:lyalpha}
\end{figure} 

The resolution of our {\it HST} STIS FUV data is not high enough to reconstruct the Ly$\alpha$ profile, as is possible with COS data \citep[e.g.,][]{schindhelm12}. However, we can check for lines produced by Ly$\alpha$ in our STIS spectra. 
If accretion-generated Ly$\alpha$ photons are responsible for the \h2 bump, it follows that when there is a change in \mdot, there is a change in Ly$\alpha$ photons, which then leads to a change in both the \h2 bump and the strength of Ly$\alpha$-driven lines.  This is not seen in our GM~Aur data.
Here we compare the brightest expected Ly$\alpha$-driven lines noted by \citet{france17} in GM~Aur between E7 and E3, the epochs with the highest and lowest \mdot, respectively. Figure~\ref{fig:lyalpha} shows no significant change in the Ly$\alpha$-driven lines. However, the \h2 bump luminosity does change between E7 and E3 (Table~5).  \citet{ingleby15} also saw no correlation between Ly$\alpha$-driven lines and \mdot using the same {\it HST} data from GM Aur E1, E2, and E3 as used in this work as well as two additional {\it HST} archival spectra from 2003 (Program 9374; PI: E. Bergin) and 2010 (Program 11616; PI: G. Herczeg).  \citet{ingleby15} suggested that this was evidence that Ly$\alpha$ is not created in the accretion shock. We do not find evidence that the \h2 bump is driven by Ly$\alpha$ photons or that Ly$\alpha$ is generated in the accretion shock.

\subsubsection{An Alternative Mechanism: The Disk Surface Density}\label{subalt}

We find that there is a strong correlation between the \h2 bump luminosity and L$_{acc}$ in our sample (Figure~\ref{fig:h2lacc}) and that the \h2 bump is much higher in GM~Aur E7 when \mdot\ was the highest (Figure~\ref{fig:h2zoom}). \citet{ingleby09} also found a clear correlation in their study of 32 CTTS with \textit{HST} STIS and ACS spectra. For the 13 objects in their sample with STIS spectra, L$_{acc}$ measurements were taken within an hour of the \h2 bump measurements (see Section 2.4); L$_{acc}$ measurements for objects in their sample with  ACS data were from the literature and therefore not coordinated in time.  \citet{france17} reported a weaker but positive correlation between the \h2 bump and \mdot\ for 24 objects where the \h2 bump was detected. For the majority of the sample, this was not based on coordinated data, and one can speculate that since \mdot\ is variable, this weakens the correlation seen by \citet{france17}. In our work, we have mostly simultaneous data and see a positive correlation between the \h2 bump luminosity and L$_{acc}$. 

The cause of the variability in L$_{acc}$ could be inhomogeneities in the inner disk that propagate through the accretion column \citep[][]{robinson17}. 
This is supported in work by \citet[][]{ingleby15} that had NIR data taken within a day of the {\it HST} data for GM~Aur E1 to E3; they found that both the \h2 bump luminosity and the dust mass in the inner disk (measured from dust continuum modeling of the NIR data) decreased by the same factor while \mdot\ decreased as well.  
Given that a decrease was seen in the NIR emission tracing the dust content of the inner disk while \mdot\ and the \h2 bump luminosity decreased as well, it is plausible this is indicating a decrease of the overall surface density in the inner disk.  Changes in the surface density in the inner disk may explain the correlation between L$_{acc}$ and the \h2 bump luminosity seen in this work.

\citet{france17} proposed that the \h2 bump is indirectly correlated to \mdot\ since the Ly$\alpha$ flux is driven by the accretion shock. However, \mdot\ also traces the surface density in the inner disk.  Therefore, it is not clear if the increase in the \h2 bump luminosity is due to a higher surface density in the inner disk or to more Ly$\alpha$ photons generated in the accretion shock. We attempt to distinguish between these two scenarios in our data by looking for correlations between Ly$\alpha$ emission lines, \mdot, and the \h2 bump, which, as noted above, we do not find (Figure~\ref{fig:lyalpha}).  
Future work could obtain coordinated {\it HST} COS and STIS data in order to measure Ly$\alpha$-driven \h2 emission lines and the \h2 bump at high resolution with COS while acquiring a more direct measure of \mdot\ and L$_{acc}$ with STIS.

\subsection{On the Role of X-ray Emission in CTTS} \label{sec:dis-xray}
 
\subsubsection{Soft X-ray Emission and the Accretion Column}

Most of the X-ray emission from CTTS has been attributed to hot, low-density (T $>10$~MK, N$_e<10^{10}$ cm$^{−3}$) plasma from coronal emission. However, there is evidence of soft X-ray emission produced by cooler (T $\sim2$--4~MK), high-density (N$_e\sim10^{12}$--$10^{13}$ cm$^{-3}$) plasma \citep{kastner02, stelzer04, schmitt05, gunther06, argiroffi07, huenemoerder07, argiroffi11}. This soft X-ray emission is seen in a few CTTS but not in WTTS \citep[e.g.,][]{telleschi07}, and so it has been attributed to accretion-related processes. Some works suggest that the high densities indicate this soft X-ray emission is formed in the postshock region at the base of the accretion column \citep[e.g.,][]{kastner02, argiroffi17}. 

However, some observations do not support the interpretation that the soft X-ray emission is formed in the accretion shock. In some objects, soft X-ray emission is present, but the plasma has lower electron densities \citep[e.g., T Tau, AB Aur;][]{gudel07c,telleschi07b} than expected if originating in the accretion shock, which should lead to higher densities than the stellar corona.  
In one case \citep[e.g., DG Tau;][]{schneider08}, the soft X-ray component has been spatially separated from the hard X-ray component; these components have been associated with the location of the jet of DG Tau and the star itself, respectively.  Also, \citet{brickhouse10} could not reproduce the densities and temperatures measured from high-resolution X-ray spectra of TW~Hya with a model of plasma heated by the accretion shock. In addition, no correlation has been found between the soft X-ray excess and UV lines known to be accretion indicators \citep{gudel07d}.
An alternative is that the soft X-ray excess is coronal plasma that is modified by the accretion process \citep{gudel07d,brickhouse10,dupree12}.

In our work, we do not see a correlation between the soft X-ray emission and L$_{acc}$.  For GM~Aur E6, E7, and E8, we have {\it Chandra} spectra that trace the soft X-ray wavelengths.  While there was a large increase in L$_{acc}$ in E7, the soft X-ray emission remained roughly constant throughout the three epochs (Figure~\ref{fig:xspec}).  This is consistent with predictions that due to the high column densities, the accretion shock is buried in the stellar photosphere and X-ray emission does not escape \citep{drake05}.
On the other hand, X-ray emission from the accretion shock is expected in some cases, particularly where the column has a lower density and high velocity \citep{sacco10}.
The change in \mdot\ in E7 may not be large enough to lead to an observable change in the continuum of the soft X-ray emission and instead high-resolution X-ray spectra would be necessary to resolve soft X-ray spectral features attributed to the accretion shock.

\subsubsection{Stellar Flares and the Accretion Column}

We find a correlation between the weighted accretion column energy flux, ${\curf}_w$, and L$_X$ (Figure~\ref{fig:curf}).  This correlation is largely driven by GM~Aur E5, which had the highest ${\curf}_w$ and L$_X$ in our sample.    
The energy flux ($\curf=0.5{\rho} v_s^3$) is proportional to the density, $\rho$, and the infall velocity, $v_s$. In our modeling, we keep $v_s$ fixed at the free-fall velocity.  This suggests a correlation between L$_X$ and the density of the accretion column. GM~Aur E5 also had the smallest sum of the filling factor, $f$, for all columns, $f_{tot}$ (= 0.06; i.e., the smallest percentage of its surface covered by accretion columns; RE19), which is consistent with narrower, denser columns.

The large increases in L$_{X}$ seen in our sample are likely due to X-ray flares from the stellar corona.
Most of the increase in the X-ray emission of GM~Aur E5 was in the hard X-ray band (Figure~\ref{fig:flares}). The GM~Aur E5 spectrum exhibits a prominent hard tail that is consistent with stellar coronal flaring activity \citep{caramazza07}. Flaring activity is also supported by the higher temperature found in spectral fitting (Table~\ref{tab:xrayflux}). 

The correlation between ${\curf}_w$ and L$_X$ may point to a connection between the accretion column and stellar flares.
Some theoretical work has found that stellar flaring activity may trigger accretion onto stars.  When flaring activity increases, there are more magnetic field lines that link the star to the disk and trigger accretion funnels onto the star \citep{orlando11,colombo19}. The predicted timescale for \mdot\ to increase after a flare ranges from a few hours to about one day and then the accretion columns themselves last between a few hours to tens of hours \citep{orlando11,colombo19}.  As the material in the accretion funnel approaches the star, the density increases due to gas compression by the dipolar magnetic field \citep{orlando11}. This is consistent with our inference of a narrower, denser column in E5 of GM~Aur.  However, we do not see an increase in \mdot. This may be attributed to the complexities of mass loading, our viewing angle, or the increase in \mdot\ occurring after our observations. Regardless, this connection between the X-ray emission and accretion column properties indicates that more observations to further explore the relationship between stellar flaring and the accretion column density would be fruitful.

\section{Summary} 

Using multiple epochs of mostly simultaneous {\it Swift/Chandra} and {\it HST} data of TTS, we found that the luminosity of the FUV \h2 bump correlates with L$_{acc}$ and not L$_X$. One mechanism to form the \h2 bump involves collisional excitation by X-ray photons.  Specifically, an increase in X-ray emission increases the ionization of the inner disk, which in turn leads to more collisional excitation of \h2.  However, we do not see evidence of a correlation between the \h2 bump and L$_X$. Another mechanism to form the \h2 bump involves Ly$\alpha$-driven dissociation of \h2O in the inner disk. A correlation between the \h2 bump luminosity and L$_{acc}$ is consistent with this scenario since the accretion funnel flow is thought to produce Ly$\alpha$ photons. However, we do not see changes in Ly$\alpha$-driven \h2 emission lines between observations where L$_{acc}$ and the \h2 bump did change significantly. Given that \mdot\ is linked to the surface density in the inner disk, we conclude that the correlation of the \h2 bump with L$_{acc}$ points to an increase in the surface density of gas in the inner disk. 

We found no correlation between L$_X$ and L$_{acc}$ or \mdot.  In addition, we do not see any changes in the soft X-ray emission in three epochs of {\it Chandra} data of GM~Aur, while \mdot\ changed by a factor of $\sim3.5$. This may support that most of the X-ray emission generated by the accretion shock is absorbed.  However, high-resolution X-ray spectra would be necessary to explore this further. We also find no correlations between L$_X$ and several FUV and NUV lines, including $C_{IV}$ and $He_{II}$.

We do see a correlation between the energy flux of the accretion columns, ${\curf}_w$, and L$_X$. This trend is dominated by coronal flaring activity. Since ${\curf}_w$ traces the density of the accretion column, this may indicate that flaring activity influences accretion onto stars. In particular, we may be seeing evidence that stellar flaring increases the amount of material lifted off the disk and onto the star, and as the material in the accretion funnel approaches the star, the density increases due to gas compression by the dipolar magnetic field.  However, we do not see an increase in \mdot, which may be due to our viewing angle or time sampling.

In conclusion, our work finds that there is no connection between the X-ray radiation field and the FUV \h2 bump in TTS. Therefore, we have yet to identify an observable tracer of the effect of X-ray ionization in the innermost disk. Instead, we find evidence that inhomogeneities in the surface density of the inner gas disk, traced by the FUV \h2 bump, propagate through the accretion column as reflected by an increase in \mdot. We also find that stellar flares may alter the accretion column density. Further coordinated multiwavelength work is necessary to understand the connection between inhomogeneities in the inner disk, X-ray emission, and mass accretion in TTS.

\appendix  

Mass accretion rates for CS~Cha and SZ~Cha are measured here following the same methods as RE19.  Stellar parameters adopted for CS~Cha and SZ~Cha are listed in Table~\ref{tab:starparam} and are taken from \citet{manara14} and scaled using new \textit{Gaia} distances \citep{gaia16,gaia18b}.
For CS~Cha, we adopt an A$_{V}$ of 0.8, a stellar radius of $R_{\ast}=1.83\;\rsun$, and a stellar mass of $M_{\ast}=1.32\;\msun$. 
For SZ~Cha, we adopt an A$_{V}$ of 1.3, a stellar radius of $R_{\ast}=1.78\;\rsun$, and a stellar mass of $M_{\ast}=1.22\;\msun$. 
We deredden the {\it HST} FUV and NUV data using the extinction law toward HD 29647 \citep{whittet04}.  We deredden the {\it HST} optical and IR data with this A$_{V}$ and the \citet{mathis90} extinction law using an R$_{V}$ of 3.1. 

To briefly summarize the methods of RE19, we follow \citet{ingleby13} and use the accretion shock models of \citet{calvet98} with multiple accretion columns.  The stellar mass, radius, and temperature are input model parameters. The accretion columns are calculated for a variety of energy fluxes, $\curf=0.5\rho v_s^3$. The energy flux depends on the density of material in the accretion column, $\rho$, and the infall velocity, $v_s$. The infall velocity is fixed at the free-fall velocity from $\sim5$ $R_{\ast}$ under the assumption that the magnetospheric radius is not changing. The resulting emission is scaled by filling factors, $f_i$, which measure the fraction of the visible stellar surface covered by the column. Finally, we calculate \mdot\ by adding the contributions of the columns with
 \begin{equation}
\mdot=\frac{8\pi R_{\ast}^2}{v_s^2} \sum_i \curf_i f_i =
\frac{8\pi R_{\ast}^2}{v_s^2} \curf_w f_{total}~~. 
 \end{equation}

As our template stellar photosphere, we adopt the WTTS RECX~1, which is in the $\eta$ Chamaeleon star-forming region. RECX~1 has STIS archival spectra obtained as part of {\it HST} proposal ID 11616 (PI: G. Herczeg), a measured A$_{V}$ of 0, and a spectral type of K5 \citep{luhman04b}. To account for the photospheric emission in order to extract the excess emission due to the accretion shock, we scale the spectrum of our WTTS template to our CS~Cha and SZ~Cha spectra. To do this properly, we must account for veiling by the excess continuum, which here we assume is due to the accretion shock. Veiling occurs when an excess continuum ``fills in'' absorption lines, causing them to appear shallower than the spectrum of a standard star of the same spectral type \citep{hartigan91}. The veiling (taken to be at 5500~{\AA}) is $r_V=F_{V,Veil}/F_{V,WTTS}$ where $F_{V,Veil}$ is the flux of the veiling continuum and $F_{V,WTTS}$ is the continuum flux of the WTTS.  
The veiling may change with \mdot\ since it has been shown that there may be some excess at optical wavelengths from accretion \citep{gullbring00, fischer11}. We cannot measure veilings from our low-resolution STIS optical spectra, so here we include them as a free parameter in our analysis. 

\begin{figure}
\epsscale{1.0}
\plotone{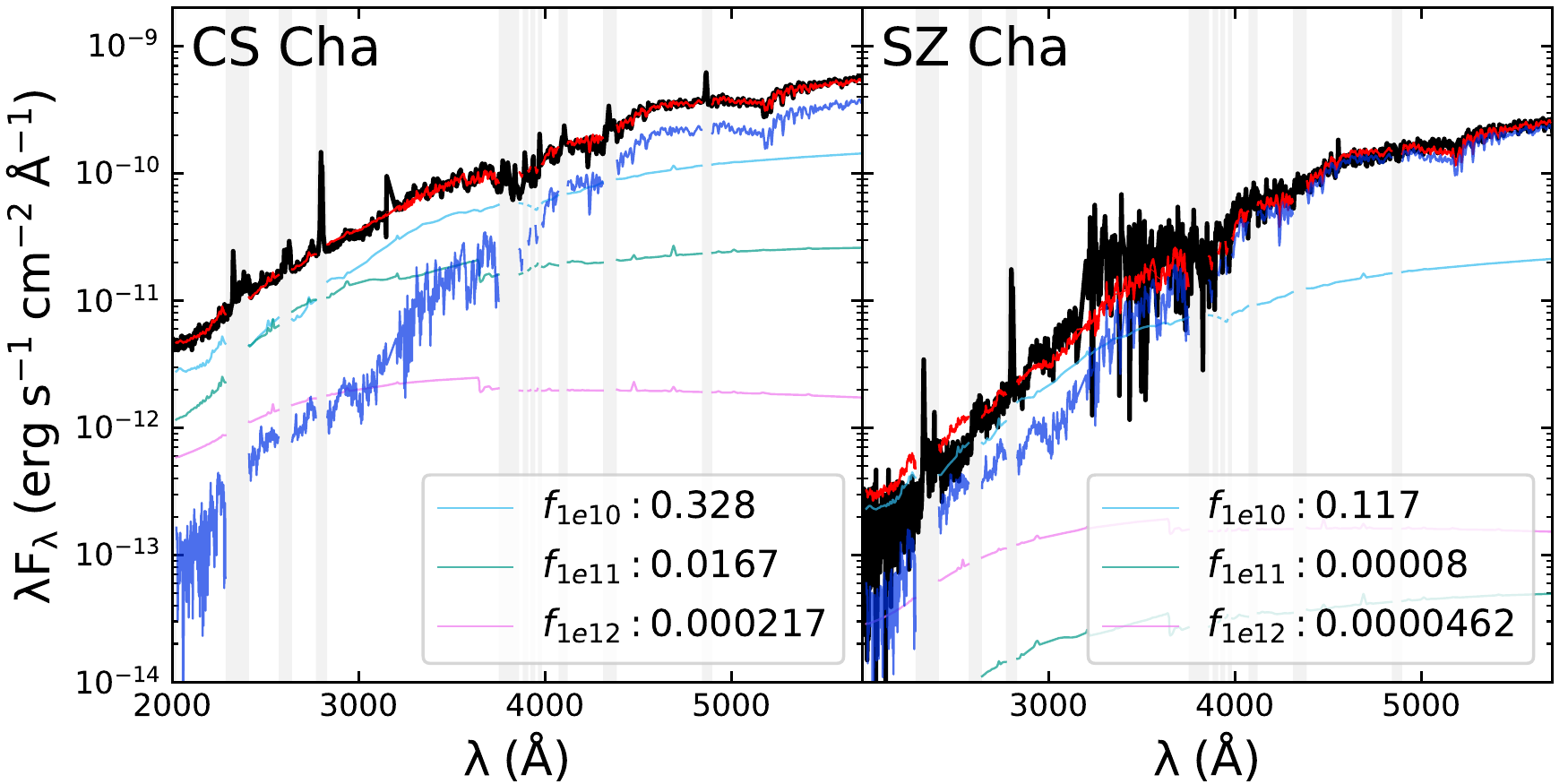}
\caption[]{
Accretion shock model fits to CS~Cha (left) and SZ~Cha (right). In each panel, we show the median total model (red) compared with the STIS spectra (black). The total model is comprised of the combined emission from three different column energy fluxes (see key) along with the undisturbed photospheric emission, which here is represented with the WTTS template RECX 1 (blue). Parameters for the best-fitting models are listed in Table~\ref{tab:mdot}. We note that here we focus on fitting the continuum emission and do not attempt to reproduce the emission lines. The gray shaded regions denote spectral features that were not included in our fitting.
}
\label{fig:mdot}
\end{figure} 

We calculated accretion shock models with $1\times10^{10}$, $1\times10^{11}$, and $1\times10^{12}$ erg s$^{-1}$ cm$^{-2}$. To model our spectra, we combined the veiled WTTS emission with the accretion column emission and left $f_i$ and $r_V$ as free parameters. The fractional uncertainty in the model is also included as a nuisance parameter. To fit these parameters, we used a Bayesian MCMC approach using the ensemble sampler \textit{emcee} \citep{foreman-mackey13}. Each parameter was fit in log-space, which eliminates the possibility of negative values. A step-function prior excludes the nonphysical cases of $f_i>1$ and $r_V<0$. An additional Gaussian prior based on previous modeling efforts by \citet{manara14} was placed on $r_V$. In Figure~\ref{fig:mdot}, we show the median model from our analysis for each target. The median values for \mdot, the filling factor per accretion shock column, and the veiling are listed in Table~\ref{tab:mdot}.

\begin{deluxetable*}{lccccc}[t]
\tablewidth{0pt}
\tablecaption{Results from Multi-Component Accretion Model Fits to {\it HST} Spectra
\label{tab:mdot}}
\tablehead{
\colhead{Target} 
&\colhead{$\mdot\; (10^{-8}\msunyr)$}
&\colhead{$f_{1E10}$}
&\colhead{$f_{1E11}$}
&\colhead{$f_{1E12}$}
&\colhead{$r_{V}$}}
\startdata
CS~Cha &
$1.497^{+0.010}_{-0.009}$ &
$0.328^{+0.007}_{-0.006}$ &
$0.0167^{+0.0006}_{-0.0006}$ &
$0.000217^{+0.000027}_{-0.000026}$ &
$0.072^{+0.008}_{-0.007}$ \\
SZ~Cha &
$0.354^{+0.011}_{-0.009}$ &
$0.117^{+0.004}_{-0.003}$ &
$0.00008^{+0.00012}_{-0.00003}$  &
$0.0000462^{+0.0000016}_{-0.0000006}$ &
$0.005^{+0.009}_{-0.004}$ 
\enddata
\tablecomments{Accretion columns with energy fluxes ($\curf$) of 1$\times10^{10}$, 1$\times10^{11}$, and 1$\times10^{12}$ erg s$^{-1}$ cm$^{-3}$ were calculated and scaled by filling factors ($f_i$).  Here we list the median values for the accretion rate (\mdot), the filling factor for each accretion column, and the veiling factor ($r_{V}$).
The corresponding positive and negative uncertainties listed are the difference between the median and the 16th and 84th percentile, respectively. We note that this is similar to 1$\sigma$ uncertainties.
}
\end{deluxetable*}

 \acknowledgments{
This work was supported by {\it HST} grants GO-13775, GO-14048, GO-14193, and GO-15165 as well as {\it Chandra} grant SAO GO8-19016A and the Sloan Foundation.
We thank the reviewer for a timely and helpful report. We thank G.~Herczeg for insightful and constructive comments that greatly improved this manuscript. We are grateful to L. Ingleby for help in planning the {\it HST} observations. We greatly appreciate valuable discussions with E.~Bergin, N.~Calvet, J.~Kastner, and D. Principe.
}

\section{}    

\bibliographystyle{aasjournal}
\bibliography{bib_2019_01}

\end{document}